\documentclass[
 reprint,
superscriptaddress,
 amsmath,amssymb,
 aps,
pra,
]{revtex4-2}

\usepackage{lineno}
\usepackage{braket}
\usepackage{graphicx}
\usepackage{dcolumn}
\usepackage{bm}
\usepackage{hyperref}
\usepackage{soul}
\usepackage{braket}

\hypersetup{colorlinks=true, citecolor=blue, urlcolor=blue, linkcolor=blue}

\begin{document}

\preprint{APS/123-QED}

\title{Approaching the Key Rate Limit in \\ Continuous-Variable Quantum Key Distribution Network}


\author{Yiming Bian}
 \affiliation{%
 State Key Laboratory of Information Photonics and Optical Communications, School of Electronic Engineering, Beijing University of Posts and Telecommunications, Beijing, 100876, China
}%
\author{Yichen Zhang}%
 \email{zhangyc@bupt.edu.cn}
 \affiliation{%
 State Key Laboratory of Information Photonics and Optical Communications, School of Electronic Engineering, Beijing University of Posts and Telecommunications, Beijing, 100876, China
}%
\author{Song Yu}%
 \affiliation{%
 State Key Laboratory of Information Photonics and Optical Communications, School of Electronic Engineering, Beijing University of Posts and Telecommunications, Beijing, 100876, China
}%
\author{Zhengyu Li}%
 \email{lizhengyu2@huawei.com}
 \affiliation{%
 Huawei Technologies Co., Ltd., Shenzhen 518129, China
}%
\author{Hong Guo}%
 \email{hongguo@pku.edu.cn}
 \affiliation{%
 State Key Laboratory of Advanced Optical Communication Systems and Networks, School of Electronics, and Center for Quantum Information Technology, Peking University, Beijing 100871, China
}%

\date{\today}

\begin{abstract}
    A quantum key distribution network enables pairs of users to generate independent secret keys by leveraging the principles of quantum physics. For end-to-end secure communication, a user pair's secret key must remain secure against any third parties, including both external eavesdroppers and other network users. However, isolating a given user pair from the remaining users while maintaining a high key rate is challenging when all users are intrinsically coupled and correlated, particularly in continuous-variable networks. This results in either a low key rate, or incomplete end-to-end security. Here, we introduce a multi-user security framework, offering a general and comprehensive end-to-end key rate formula, against the collaboration of the other network users and eavesdropper. Building on this framework, we propose a multi-user protocol that achieves the theoretical upper limit in all practical deployments within a 100 km range. Applied to a three-node network, it achieves an Mbps-level per-user key rate and an overall network key rate reaching 90\% of the upper limit. The proposed solution supports scalable implementations using telecom-compatible components, while the method for obtaining the accessible information in the network is broadly applicable and can be extended to various multipartite quantum information systems. 

\end{abstract}

\maketitle

Quantum key distribution (QKD) enables a pair of users, Alice and Bob, to generate shared random secret key bits against a potential eavesdropper (Eve) with information-theoretic security \cite{bennet1984quantum, AdvInQC, PTPQKDRMV2020, portmann2022security}. To facilitate large-scale implementations, point-to-point QKD has been generalized to QKD networks \cite{QCnetNature1997,wang2014field,QNetScience2018,CambridgeQN,QCNetToyoko,QCnetNature2021}, allowing any two users in the network to generate secret keys against both eavesdropper and other network participants. While isolating individual user pairs is straightforward in single-photon-based QKD networks \cite{QCnetNature1997,wang2014field,QNetScience2018,CambridgeQN,QCNetToyoko,QCnetNature2021}, continuous-variable QKD (CV-QKD) \cite{GG02PRL,NSPRL,GaussianQuantumInformation} can inherently correlate all participants, particularly when multiple users share the same network server \cite{Huang2021Realizing,bian2023high,hajomer2024continuous,pan2024high}. In such CV-QKD networks, the remaining participants are treated as adversaries, as they gain access to the information about the user generating secret keys. The overhead required to eliminate this residual information scales with the network capacity, leading to a significant degradation in secret key rates.

Until now, various multi-user CV-QKD protocols have been proposed to address the coupled users and enhance the key rate \cite{Huang2021Realizing,bian2023high,hajomer2024continuous,pan2024high}. However, perfect user isolation and a high key rate cannot be achieved simultaneously. One approach prioritizes user isolation by assuming that the remaining legitimate participants are fully controlled by Eve \cite{Huang2021Realizing}, leading to an overly pessimistic key rate. Conversely, other protocols achieve a high key rate but rely on the trusted user assumption, where Eve is assumed to have no access to any raw-key information from some \cite{hajomer2024continuous} or all \cite{bian2023high,pan2024high} of the remaining participants. This leads to incomplete end-to-end security, as the security of a given user pair depends on all network participants. Any other user's violation introduces a side channel, compromising the security of the entire network. Moreover, the maximum key rate achievable by the network remains unresolved, leaving the ultimate key rate and the fundamental benchmark for CV-QKD networks unclear.

In this Letter, we establish an explicit end-to-end key rate formula for general multi-user CV-QKD systems, along with the lower and upper key rate limits. Eve's benefit from the remaining users is precisely quantified, resulting in an accurate estimation of the accessible information. Within this framework, we propose a multi-user CV-QKD protocol that achieves the theoretical upper limit in all practical deployments within a 100 km range. It is then applied to a 3-node CV-QKD network, resulting in an Mbps-level secret key generation rate per user, and an overall network key rate achieving 90\% of the theoretical upper limit. The results not only demonstrate the effectiveness of the proposed security analysis tools but also reveal that high key rates and comprehensive end-to-end security can be simultaneously achieved in CV-QKD networks without complex physical isolation of end users. Furthermore, telecom compatibility of CV-QKD ensures the scalability \cite{Bian2024Continuous,hajomer2024continuous10GBaud,pietri2024experimental,Zhang2024Continuous}, and the explicit calculation of the network accessible information extends to general multipartite quantum information systems.

\begin{figure}
    \centering
    \includegraphics[width= 7.5 cm]{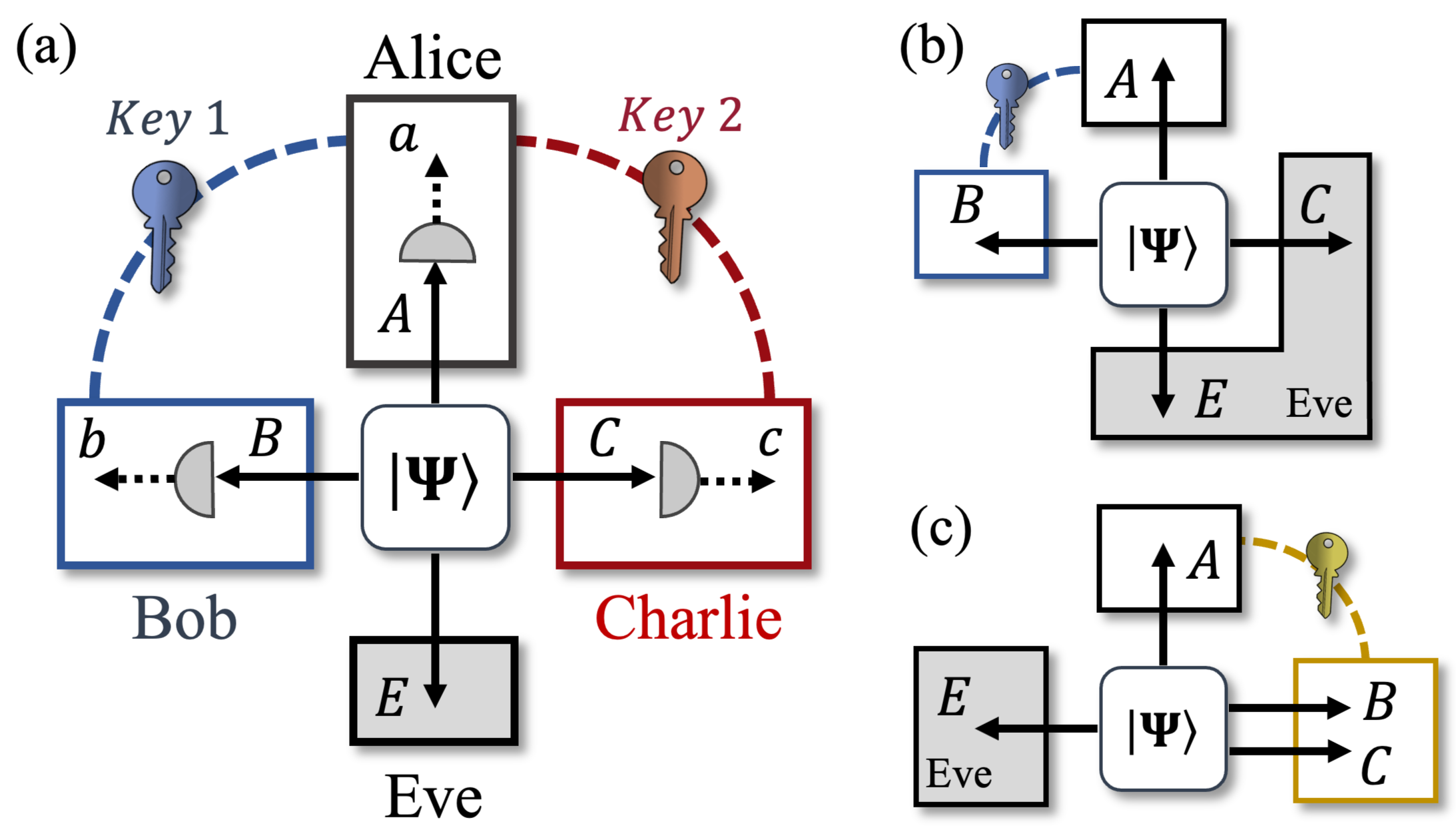}
    \caption{(a) A 3-node CV-QKD network. Alice generates independent secret keys with Bob and Charlie based on state $\rho_{ABC}$. Eve can purify the whole network ($ABC$), therefore, the overall system ($ABCE$) is a pure state ($\ket{\Psi}$). (b) The worst-case scenario, assuming that Eve fully holds the rest of the network. (c) The best-case scenario, where Bob and Charlie are trusted and located in the same site. Here, $a$, $b$ and $c$ represents the measurement results of $A$, $B$ and $C$.
    }\label{Fig: Network}
\end{figure}

\textit{Multi-user QKD---}
We first consider a general three-user quantum network consisting of the legitimate participants Alice ($A$), Bob ($B$), and Charlie ($C$). They share a quantum state $\rho_{ABC}$, as illustrated in Fig. \ref{Fig: Network} (a). The eavesdropper Eve can purify the entire system $ABC$, and her purification is denoted as $E$. Each legitimate participant performs measurements on their respective subsystems, obtaining the outcomes $a$, $b$, and $c$. For comprehensive end-to-end security, any user pairs' secret key must remain secure even if Eve can access other users' secret information. Considering the Alice-Bob pair, the required key rate is given by $K_{AB} = \beta I(a:b) - \chi(b:E\tilde{c})$, where $\tilde{c}$ represents Eve's benefit from Charlie, and $\beta$ is the reconciliation efficiency. 

In the worst case, Eve is assumed to have full control over Charlie, i.e., $\tilde{c} = C$. The corresponding network scheme is shown in Fig. \ref{Fig: Network} (b). A key rate estimated under such conditions guarantees the security against both Eve and the remaining network participants. However, it is overly pessimistic, as mode $C$ is detected by Charlie rather than manipulated by Eve. Such assumption maximizes Eve's ability and sets the lower bound of the key rate, i.e., $K_{LB} = \beta I(a:b) - \chi(b:EC)$. Conversely, the best-case scenario, depicted in Fig. \ref{Fig: Network} (c), assumes that Bob and Charlie are trusted and co-located. Alice can generate secret keys with the collective mode $BC$. It transforms the multi-user system to a point-to-point system, avoiding Eve's access to Charlie's information and offering an upper limit $K_{UB} = \beta I(a:bc) - \chi(bc:E)$.

In practical network implementations, Bob and Charlie are typically located at separate sites. Despite physically isolating the user pairs, i.e., $\rho_{ABC} = \rho_{A_1 B} \otimes \rho_{A_2 C}$, networks with inherently coupled users inevitably face a trade-off between achieving high key rate and ensuring complete end-to-end security, particularly in CV-QKD networks. To achieve high key rate, the network relies on trusted users, assuming that Eve can never access Charlie's (or Bob's) information when Alice generates secret keys with Bob (or Charlie). Specifically, this requires that Eve cannot obtain the measurement results $b$ and $c$, nor any data transformed from the measurement results, e.g., the syndromes $s(b)$ and $s(c)$. 

To avoid the information leakage caused by syndromes, two protocols have been proposed, relying on masking the syndromes \cite{pan2024high} or establishing a hierarchical system of trust among users \cite{hajomer2024continuous}. However, the assumption that the other network participants' measurement results remain completely inaccessible to Eve is still required. Despite the high key rates, the coupled users and their correlations make the existing protocols fragile in several ways: (1) Eve can attack the Alice-Bob link without directly accessing Alice's or Bob's site. She can target Charlie and gain access to $c$, which breaks the trusted-user assumption and creates a security loophole. It leads to incomplete end-to-end security. (2) Masking syndromes requires additional operations, increasing the network complexity and making it susceptible to reconciliation failures, which can interrupt the protocol. (3) The hierarchical trust approach relies on the assumption that the syndromes disclosed by users who have completed key generation remain secure. Compromising of this assumption reopens the user-correlation-induced side channel.



\textit{End-to-end key rate---}
Given that mode $C$ is measured, Eve's only potential way to maximize her knowledge about mode $b$ is by completely obtaining Charlie's measurement result $c$ and collectively processing the system $Ec$, i.e., $\tilde{c} = c$. Note that $s(c)$ is the local function of $c$, thus, if Eve already holds $c$, obtaining $s(c)$ provides her no additional information. Alice-Bob's end-to-end key rate is thereby given by $K_{AB} = \beta I(a:b)-S(b:Ec)$. Similarly, the key rate between Alice and Charlie is $K_{AC} = \beta I(a:c)-S(c:Eb)$. The total network key rate is therefore $K_{tot} = K_{AB} + K_{AC}$. This marks a significant improvement over previous protocols, as our approach eliminates the need for trusted user assumptions and syndrome masking. By considering the collaboration of Eve and the remaining network participants, it removes all side channels caused by user correlations without introducing additional assumptions or complexity to the network.

The key to achieving a high secret key rate lies in the tight estimation of $S(b:Ec)$ (and $S(c:Eb)$) using the experimentally accessible data. Current protocols scale $S(b:Ec)$ to $S(b:EC)$, providing only a lower bound on the key rate. Here, we demonstrate that $S(b:Ec)$ can be directly computed, resulting in a simple and intuitive key rate formula. The security analysis is based on the classical-quantum state 
\begin{equation}
    \begin{aligned}
        \rho_{bcE} = \iint db\ dc\ p(b,c) \ket{b}\bra{b} \otimes \ket{c}\bra{c}\otimes \rho_{E}^{bc}.
    \end{aligned}
\end{equation}
From this, we obtain
\begin{equation}
    \begin{aligned}
        \rho_{Ec} & = \int dc\ p(c) \ket{c}\bra{c}\otimes \rho_{E}^{c},\\
        \rho_{Ec}^{b} & = \int dc\ p(c|b) \ket{c}\bra{c}\otimes \rho_{E}^{bc}.
    \end{aligned}
\end{equation}
The joint entropy theorem indicates that the Von Neumann entropy of a classical-quantum state can be decomposed into two parts: the Shannon Entropy of the classical state and the average Von Neumann entropy of the remaining system \cite{nielsen2010quantum}. Therefore, we have
\begin{equation}\label{eq:S_state}
    \begin{aligned}
        S(\rho_{Ec})&=H(p(c))+\int d{c} \ p(c)S(\rho_{E}^{c}),\\
        S(\rho_{Ec}^{b})&=H(p(c|b))+\int dc \ p(c|b)S(\rho_{E}^{bc}).
    \end{aligned}
\end{equation}

For an Eve collaborating with Charlie and having access to the classical information $c$, her knowledge about $b$ is bounded by the quantum mutual entropy coinciding with the Holevo bound ($\chi_{b:Ec}$) \cite{holevo1973bounds}, namely,
\begin{equation}\label{eq:Qmutual}
    S(b:Ec) = \chi_{b:Ec} = S(\rho_{Ec})-\int db \ p(b)S(\rho_{Ec}^{b}).
\end{equation}
It it also known as accessible information \cite{nielsen2010quantum}. Given that Eve can purify system $\rho_{ABC}$, we obtain $S(\rho^{c}_{E}) = S(\rho^{c}_{AB})$ and $S(\rho^{bc}_{E}) = S(\rho^{bc}_{A})$ when rank-1 measurements are used to get $b$ and $c$. Since $H(c) = H(p(c))$, $H(c|b) = \int db \ p(b) \ H(p(c|b))$ and $I(b:c) = H(c)-H(c|b)$, we obtain a simplified form of Eve's knowledge from Eq. \ref{eq:S_state} and Eq. \ref{eq:Qmutual},
\begin{equation}\label{eq:S0}
    \begin{aligned}
        S(b:Ec) = & I(b:c)+\int dc \ p(c) S(\rho_{AB}^c) \\ 
        & -\int db \ p(b)\int dc \ p(c|b)S(\rho_{A}^{bc}).
    \end{aligned}
\end{equation}
Eq. \ref{eq:S0} can be further simplified when $ABC$ is Gaussian and the legitimate participants use homodyne or heterodyne detection. In this case, the Von Neumann entropy of the remaining system after detection is independent to the measurement results. It indicates that $S(\rho_{AB}^c)$ and $S(\rho_{A}^{bc})$ remain the same for any $b$ and $c$. Therefore, we have
\begin{equation}\label{eq:S}
    \begin{aligned}
        S(b:Ec) = & I(b:c) + S(\rho_{AB}^c) -S(\rho_{A}^{bc})\\
        = & I(b:c) + S(b:E |c).
    \end{aligned}
\end{equation}
It is straightforward to generalize mode $C$ to $N$ modes, denoted as $\mathbf{C} = C_1 C_2 \dots C_N$, where the sub-modes are independently measured, producing outputs $\mathbf{c} = c_1 c_2 \dots c_N$. In a Gaussian system, $I(b:\mathbf{c})$, $S(\rho_{AB}^\mathbf{c})$ and $S(\rho_{A}^{b\mathbf{c}})$ can be easily calculated using the covariance matrix $\gamma_{ABC_1 C_2 ... C_N}$ estimated by experimental data, detailed in supplementary materials. Intuitively, it decomposes Eve' s knowledge into two parts when she collaborates with the remaining network participants: (1) the classical mutual information held by the remaining network participants, and (2) Eve's quantum mutual information excluding the benefits from the above classical information. In supplementary materials we prove that the Gaussian attack is optimal, thereby Eq. \ref{eq:S} is the rigorous upper limit of Eve's information.

\begin{figure}
    \centering
    \includegraphics[width= 7.5 cm]{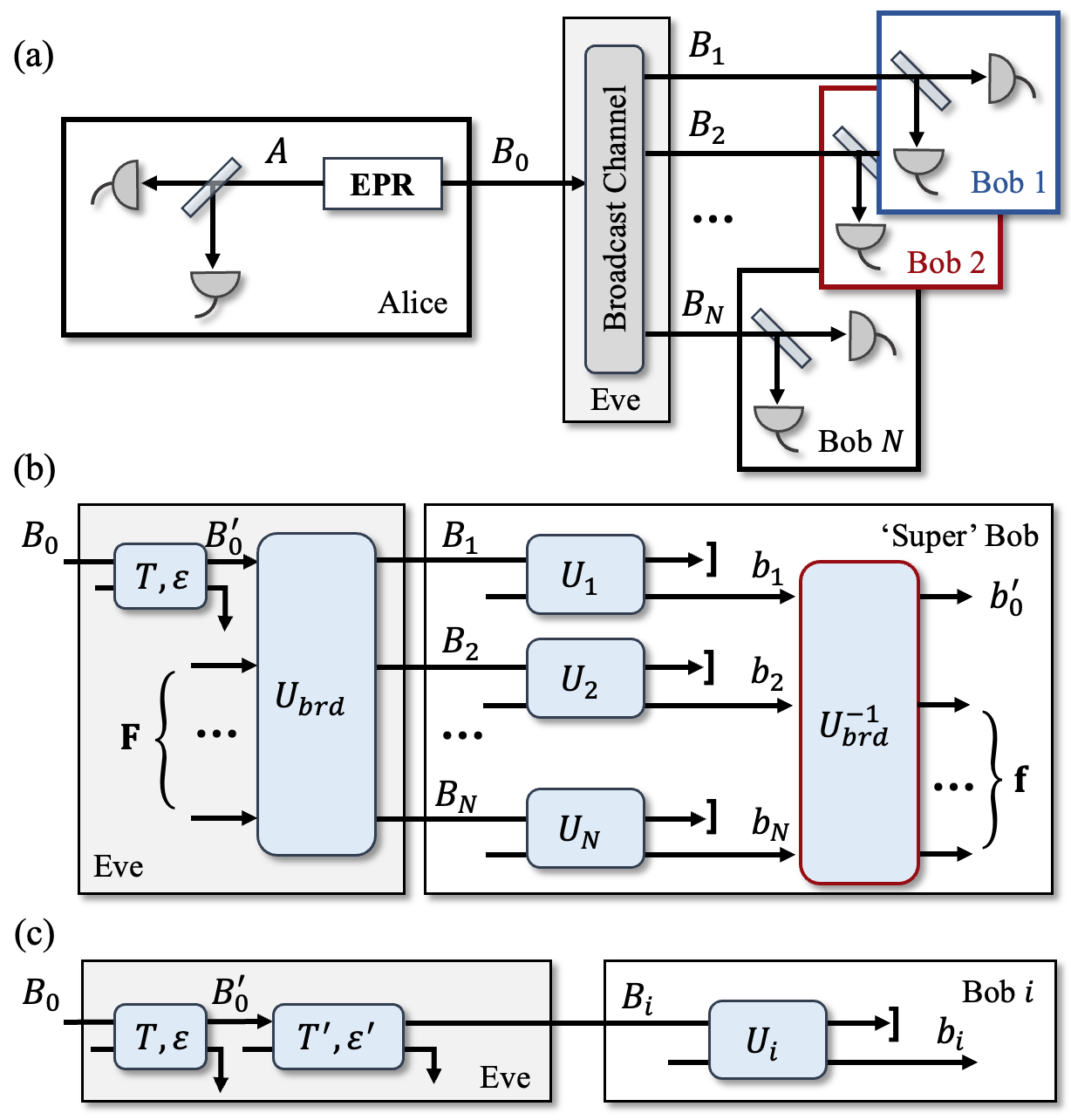}
    \caption{(a) The entanglement-based scheme of a practical multi-user CV-QKD protocol by broadcasting the mode $B_0$. Here, Eve can manipulate the broadcast channel. (b) The model of the best-case scenario. All Bobs are co-located, recovering the multi-user protocol to a point-to-point protocol by canceling the broadcast operation with an inverse transformation. (c) The model of the worst-case scenario. All Bobs except for Bob $i$ are untrusted. The broadcast operation introduces transmittance $T^{'}$ and excess noise $\varepsilon^{'}$, leading to a lower bound of the key rate.
    }\label{Fig: Protocol}
\end{figure}

\textit{Practical protocol with broadcast channel---}
The most straightforward method for establishing a multi-party state in a CV-QKD network is through source broadcasting, as illustrated in Fig. \ref{Fig: Protocol} (a). Here, Alice generates a quantum state $\rho_{AB_1B_2...B_N}$ by broadcasting one mode ($B_0$) of an EPR state to $N$ Bobs. Each legitimate party performs heterodyne detection on their respective modes, then Alice generates independent secret keys with each Bob. Eve, meanwhile, has the capability to manipulate the broadcast channel and purify the entire network. This protocol indeed corresponds to a practical multi-user CV-QKD scenario, specifically, the CV-QKD access network based on a passive optical network. It accesses massive end users to the network by generating independent secret keys between the network server (Alice) and multiple end users (Bobs). The broadcast channel can be realized using an optical fiber beam splitter, and the source preparation can be implemented by modulating the coherent states. Given by Eq. \ref{eq:S}, we obtain the key rate between Alice and Bob $i$ as
\begin{equation}\label{eq:key}
    K_{AB_i} = \beta I(a:b_i) - I(b_i:\mathbf{b_r}) - S(b_i : E|\mathbf{b_r}),
\end{equation}
with $\mathbf{b_r}$ the classical modes of the remaining Bobs (e.g., for $i=1$, $\mathbf{b_r} = b_{2,3,...,N}$). The total key rate of the network is thus given by $K_{tot}=\sum_{i}K_{AB_i}$.

The upper and lower limits of the key rate for multi-user CV-QKD with a broadcast channel are derived as follows. When Bobs are co-located and collaborate, they can jointly process the measurement results to maximize the key rate. The upper limit of the key rate is thus given by $K_{UB} = \beta I(a:b_1b_2...b_N)-S(b_1b_2...b_N:E)$. Fig. \ref{Fig: Protocol} (b) presents a model of the broadcast channel, which consists of a one-way channel characterized by transmittance $T$ and excess noise $\varepsilon$, followed by a unitary operation $U_{brd}$ acting on $B_{0}^{'}$ and ancilla modes $\mathbf{F} = F_1 F_2 ... F_{N-1}$. Each Bob's heterodyne detection is modeled by applying an appropriate unitary operation $U_i$ with an ancilla, tracing over the resulting system, and observing the output ancilla system ($b_i$) \cite{CVSecure2006second}. Using these models, the `super' Bob can reverse the broadcast operation with $U_{brd}^{-1}$. Since the measurement process can be interchanged with $U_{brd}^{-1}$, the measurement results $\{b_{0}^{'},\mathbf{f}\}$ can be recovered. These are exactly the results by directly measuring mode $B_{0}^{'}$ and $\mathbf{F}$. Given that mode $\mathbf{F}$ has no correlation with mode $A$ and $B_{0}^{'}$, and recognizing that the secret key rate remains invariant under local unitary transformations, the upper limit simplifies to $K_{UB} = \beta I(a:b_{0}^{'})-S(b_{0}^{'}:E)$. This is exactly the point-to-point key rate without broadcast operation, which coincides with the conclusion in \cite{laurenza2017general}.
Fig. \ref{Fig: Protocol} (c) illustrates the worst-case scenario, where all Bobs except for Bob $i$ are untrusted. If the broadcast operation introduces the transmittance $T^{'}$ and excess noise $\varepsilon^{'}$, the key rate lower bound, $K_{LB}$, is exactly the point-to-point key rate with transmittance $TT^{'}$ and channel input excess noise $\varepsilon + \varepsilon^{'}/T$.

\begin{figure}[t]
    \includegraphics[width=8.5 cm]{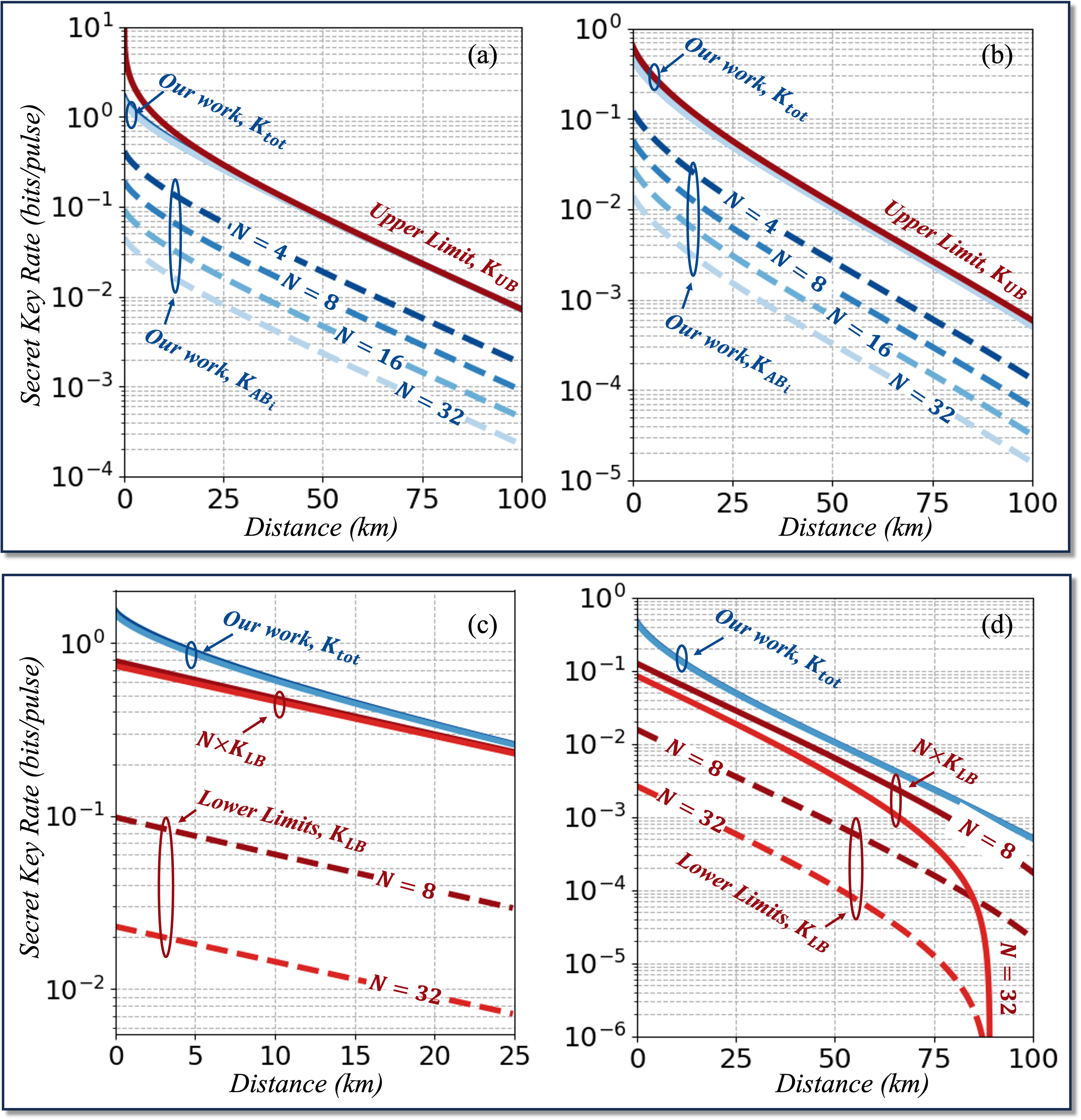}
    \caption{\label{fig:Simulation1}
    Protocol performance compared to upper (a and b) and lower (c and d) limits under ideal (a and c) and practical (b and d) conditions. (a and b) show per-user ($K_{AB_i}$) and total key rates ($K_{tot}$) for $N=$ 4, 8, 16 and 32, comparing with the upper limit ($K_{UB}$). Note the near coincidence of $K_{tot}$ curves across varying $N$. (c and d) show $K_{tot}$ curves for $N=$ 8 and 32, comparing with the lower limits, $K_{LB}$ and $N \times K_{LB}$, corresponding to the protocol in Ref. \cite{Huang2021Realizing} and \cite{hajomer2024continuous}. 
    Simulation parameters: Ideal condition with reconciliation efficiency $\beta=1$, modulation variance $V_M=10^4$ SNU and excess noise $\varepsilon=0$. Practical condition with $\beta=0.956$, $V_M=4$ SNU, $\varepsilon=0.05$ SNU and a detector with detection efficiency $\eta=0.6$, electronic noise $\nu_{ele}=0.1$ SNU.
    }
\end{figure}

\textit{Achieving the upper limit---} The secret key rates of the multi-user protocol with untrusted users are simulated and compared with the upper and lower key rate limits in Fig. \ref{fig:Simulation1}. In extreme cases where all Bobs exhibit high correlation, achieved by setting a large modulation variance, the total key rate of the protocol ($K_{tot}$) reaches the upper limit ($K_{UB}$) with a transmission distance over 25 km (see Fig. \ref{fig:Simulation1} (a)). In practical situations shown in Fig. \ref{fig:Simulation1} (b), $K_{tot}$ attains $K_{UB}$ at all transmission distances within 100 km, demonstrating the tightness of our key rate formula. This indicates that collaborating with the remaining network participants basically provides no benefits to Eve under our security analysis framework. Furthermore, though the increase of the number of Bobs introduces a higher channel loss and limits the per-user key rate, the total key rate remains stable, suggesting that the overall performance is minimally constrained by network capacity. This contrasts with the previous protocol \cite{Huang2021Realizing}, where the key rates decline significantly with an increasing number of users due to a simplistic and overly pessimistic parameter estimation.

As demonstrated in Fig. \ref{fig:Simulation1} (c) and (d), our approach significantly outperforms the key rate lower bound ($K_{LB}$) by several orders of magnitude, with performance scaling favorably with the number of users. Compared to $N \times K_{LB}$, our approach achieves superior performance across all distances in both ideal (see Fig. \ref{fig:Simulation1} (c)) and practical scenarios (see Fig. \ref{fig:Simulation1} (d)). Notably, under practical long-distance transmission with $N=32$, $K_{LB}$ and $N \times K_{LB}$ drop to zero, while our protocol still maintains a high key rate. 
The performance enhancement arises from the tight estimation of Eve's knowledge when she collaborates with the remaining network participants. This is based on the useful tool we developed for calculating the maximum mutual information between a classical party (a Bob) and a classical-quantum party (the joint collaboration of Eve and the remaining Bobs). It is worth noting that, in practical situations, the limited modulation variance and the detector imperfections weaken the correlation between Bobs, thereby eliminating Eve's advantage in getting the measurement results of the remaining network participants. As a result, the protocol key rate always achieves the upper limit while untrusting all remaining participants, demonstrating optimized multi-user CV-QKD performance with high-level practical security and a streamlined implementation.

\begin{figure*}[t]
    \includegraphics[width= 17 cm]{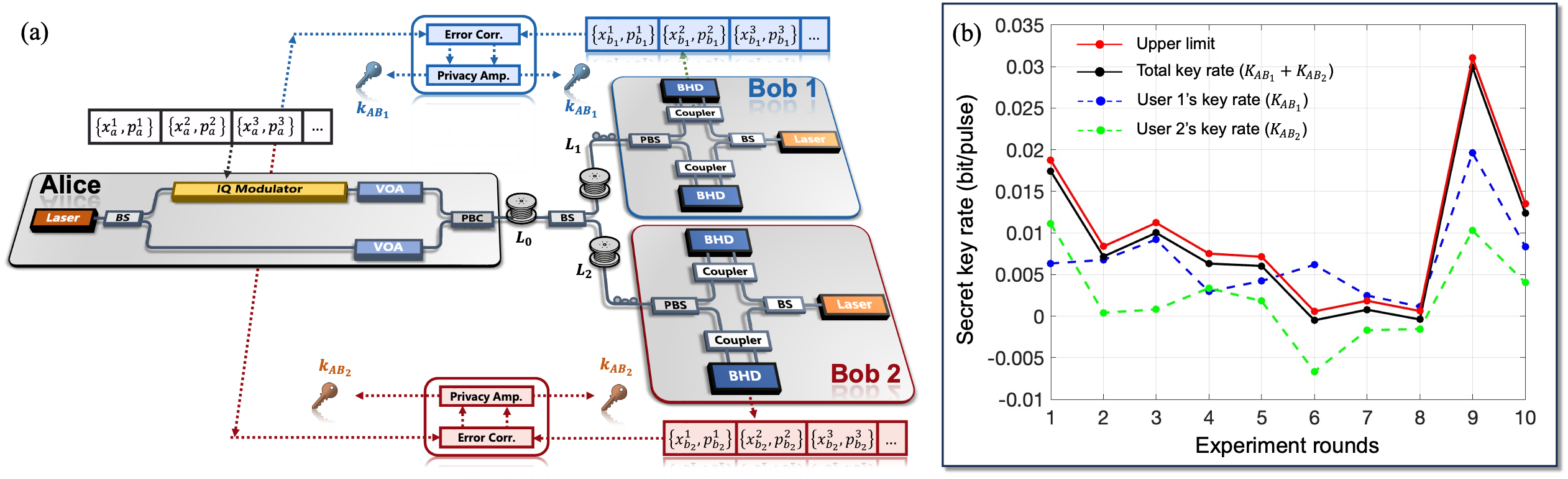}
    \caption{\label{Fig: Exp_Scheme}
    (a) The schematic of the proof-of-principle experiment. Gaussian modulation is realized by an IQ modulator and a variable optical attenuator. After the transmission through a fiber link with the length of $L_0 = 25 \ km$, the quantum signal is divided by a 1-to-2 beam splitter, then sent to two receivers with the same configuration through two fiber links with the length of $L_1 = L_2 = 5 \ km$. 
    (b) The secret key rate results in 10 rounds. User 1's and user 2's secret key rates (blue and green dashed lines), and the total key rates (black solid line) of the network are compared with the upper limit (red solid line). Here, the upper limit is calculated with the covariance matrix experimentally estimated in each round. 
    }
\end{figure*}

\textit{Experimental proof of principle---} The proposed protocol has been experimentally demonstrated with a 3-node network testbed as shown in Fig. \ref{Fig: Exp_Scheme} (a). Each coherent state produced by Alice makes the two Bobs response simultaneously. Each Bob owns an independent laser to provide the local oscillator \cite{CvExpLLO2015,CvExpLLO2015-2}. One-time shot noise unit calibration is used for simplicity and accuracy \cite{zhang2020one}. Bob 1 and Bob 2's detection efficiencies are 0.502 and 0.485. Their excess noise are 0.085 and 0.103 SNU (at the channel input). Following Eq. \ref{eq:key}, we have the average secret key rates $K_{AB_1}=6.7\times 10^{-3}$ bit/pulse and $K_{AB_2}=2.2\times 10^{-3}$ bit/pulse with a reconciliation efficiency of 96\% \cite{CvExp202kmPRL}. 
Fig. \ref{Fig: Exp_Scheme} (b) presents the two users' key rates of 10 experiment rounds, along with the total key rate ($\sum_i K_{AB_i}$) and the upper limit that the network can achieve in each round of experiment. The total key rate reaches 90\% of the upper limit in the 3-user network practical implementation, indicating a low cost for isolating different network participants. The system baud rate is 1 GBaud, and the overhead caused by parameter estimation and training symbols are 50\%. Therefore, the final secret key generation rate is 3.36 and 1.09 Mbps for user 1 and user 2, and the total key rate is 4.45 Mbps. 




\textit{Discussions---} In this work, we theoretically and experimentally demonstrate a multi-user CV-QKD that simultaneously achieves high performance, robust practical security, and straightforward implementation. This is enabled by the development of a general tool for calculating the accessible information in a multi-party quantum system, particularly when an ancilla system involves into the information distillation process. In the multi-user CV-QKD scenario we study, the rest network users act as an ancilla system, providing Eve their measurement results but cannot be fully controlled by her. Compared to existing security analysis methods assuming that the remaining users are fully controlled by Eve, our approach separates the accessible information of the ancilla system, providing a tighter estimation that results in a substantial enhancement in key rate and practical security. Moreover, the proposed approach extends beyond CV-QKD networks to a broad range of quantum information systems, enabling the decomposition of accessible information of any multi-party quantum system that can be characterized by a classical-quantum state.

In the aspect of building practical and scalable CV-QKD networks, the results show a counterintuitive conclusion: For end-to-end key generation in a CV-QKD network, holding the raw data of other correlated users provides negligible benefits to Eve. This implies that the user isolation required by current networks can be alternatively eliminated while maintaining the same or even superior key rates. It enables the construction of a CV-QKD access network that maintains the same performance while utilizing a simplified architecture and lower-bandwidth devices. Moreover, the proposed protocol extends beyond access networks, enabling novel CV-QKD network topologies (detailed in End matters). 
The proposed scheme is also compatible with non-Gaussian sources \cite{DMCVLeverrier,DMCVLinjie} and passive state preparation techniques \cite{qi2020experimental}. Additionally, it can support entanglement-based CV-QKD networks where the multi-mode state is directly prepared \cite{jia2025continuous}. Future applications of this approach across various platforms and scenarios will fully exploit the unique advantages of CV-QKD in networking, paving the way for high-performance, large-scale, and telecom-compatible quantum communication networks.


This research was supported by the National Natural Science Foundation of China (U24B20135), the National Cryptologic Science Fund of China (2025NCSF02050), and the Equipment Advance Research Field Foundation (315067206).

\bibliography{apssamp}

\onecolumngrid
\subsection*{End Matter}
\twocolumngrid

\begin{figure}[]
    \includegraphics[width= 8.5 cm]{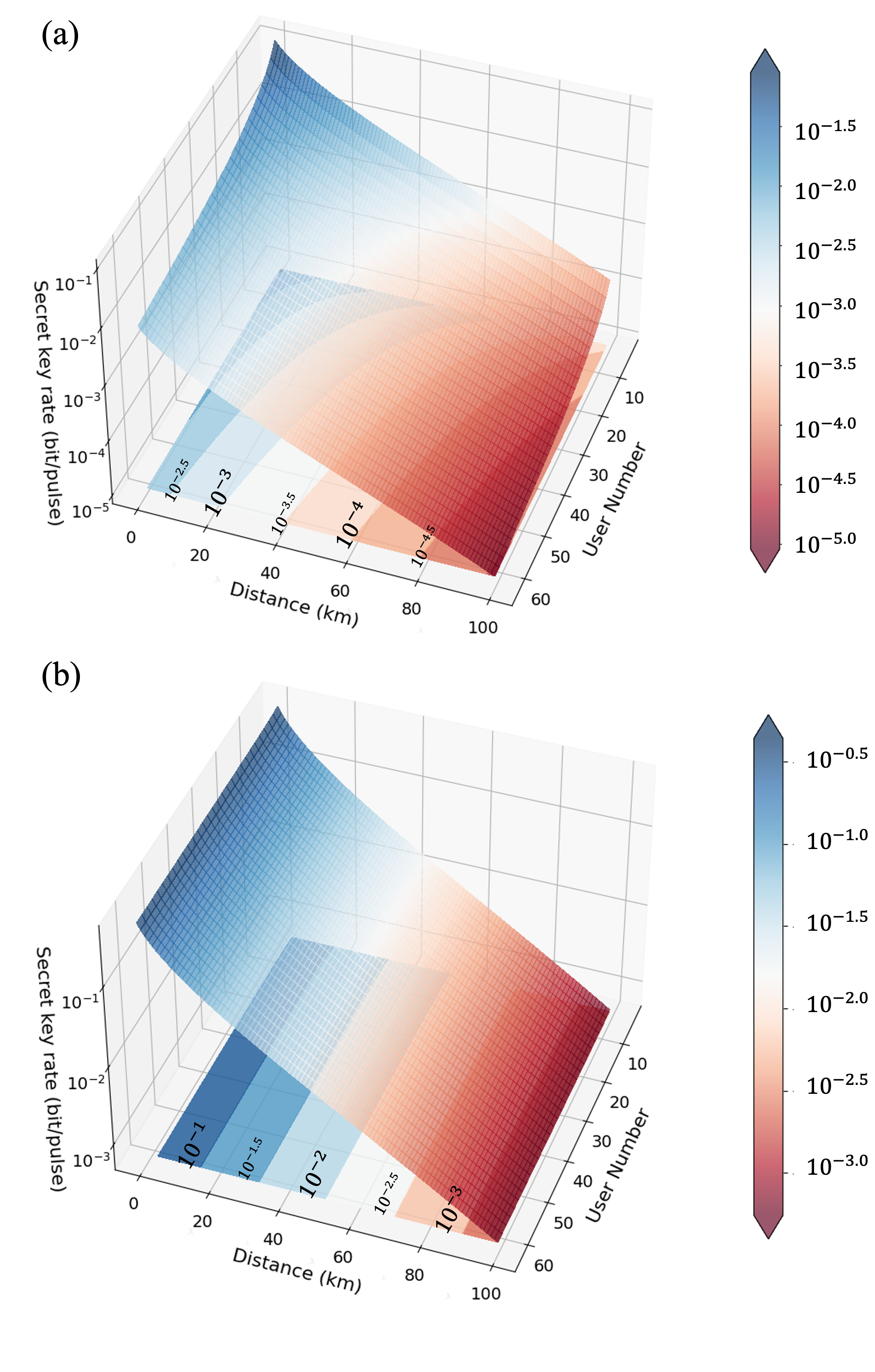}
    \caption{\label{fig:Simulation3}
    Visualization of the per-user key rate (a) and total key rate (b) in practical situation. The 3D surface illustrates the secret key rate as a function of transmission distance and the number of users. The accompanying 2D projection displays the corresponding range of secret key rates across varying distances and user numbers. The simulated user numbers are in the range: 4 to 64.  Simulation parameters: $\beta=0.956$, $V_M = 4$ SNU, $\varepsilon=0.05$ SNU, detection efficiency $\eta=0.6$, electronic noise $\nu_{ele}=0.1$ SNU.
    }
\end{figure}
\textit{Protocol performance---}
The proposed multi-user protocol achieves the upper limit and offers stable, high key rates. Fig. \ref{fig:Simulation3} (a) and (b) illustrate the per-user and the total network key rates under realistic simulation parameters. In access distances (within 20 km), the per-user key rate exceeds $10^{-3}$ bit/pulse for network capacities up to 64 users. At a typical metropolitan distance of 60 km, the protocol achieves a secret key rate exceeding $10^{-4}$ bit/pulse. While the per-user key rate reduces with the increase of the user number, the total network key rate remains stable across varying network capacities. It suggests that the untrusted channel loss incurred by broadcasting the source to a large number of end users does not degrade the total extractable information from the source.

The results demonstrate that Mbps-level quantum access network services can be delivered using only 10 MHz-class bandwidth devices. As shown in Fig. \ref{fig:Simulation3}(b), the total network key rate exceeds $10^{-1}$ bit/pulse within 10 km. Therefore, a 1-to-4 network operating at a symbol rate of 40 Mbaud can achieve a total key rate of 4 Mbps, providing 1 Mbps per user. This greatly simplifies network deployment, enabling scalable access for large numbers of quantum end users. Even at distances of 60-100 km, a total key rate of $10^{-3}$ bit/pulse is attainable, supporting Mbps-level metropolitan network services with Gbps-class devices. These results highlight the protocol's suitability for both access and metropolitan network scenarios, showing its practicality and real-world applicability.

\begin{figure}
    \centering
    \includegraphics[width= 8 cm]{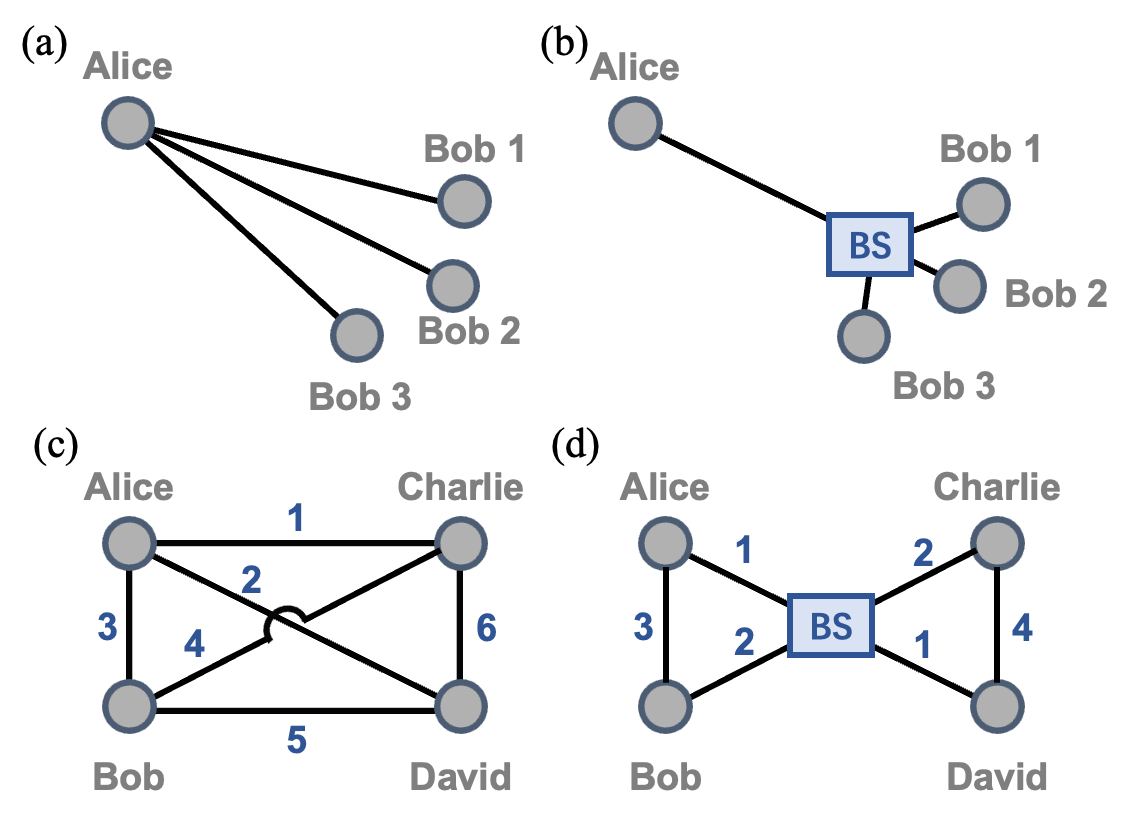}
    \caption{The CV-QKD networks. (a) A star network based on point-to-point connections. (b) A point-to-multipoint network based on broadcast channel and the proposed multi-user protocol. (c) A 4-node relay-less CV-QKD network with point-to-point connections. (d) A 4-node relay-less CV-QKD network with the proposed multi-user protocol.}\label{Fig: Discuss}
\end{figure}

\textit{Simple network topologies---}
The proposed multi-user protocol can be deployed in access and metropolitan distances, enabling relay-less high-rate CV-QKD networks with simple topology and less channel usage. 
As shown in Fig. \ref{Fig: Discuss} (a), the traditional star network requires the point-to-point connections between the network server (Alice) and end users (Bobs). This topology is normally used in access and short metropolitan distances ($\leq 25$ km) and the a crucial limitation is the `last-mile problem': Accessing a large number of end users necessitates the deployment of optical fibers on a significant scale. Assuming an average required optical fiber transmission distance of $L_0$ between Alice and each Bob $i$, then a total fiber length of $N \times L_0$ is required for the network topology depicted in Fig. \ref{Fig: Discuss} (a). 

The point-to-multipoint network topology, which utilizes a broadcast channel as illustrated in Fig. \ref{Fig: Discuss} (b), can address this challenge, especially when the Bobs are situated in close proximity, such as within the same campus or building. The broadcast channel consists of a feeder fiber of length $L_1$ and multiple drop fibers of an average length of $L_2$. The quantum state prepared by Alice is sent to the beam splitter with feeder fiber, then broadcasted to multiple Bobs with drop fibers.The proposed protocol allows all Bobs to simultaneously generate independent and secure secret keys with Alice, while maintaining a total key rate that is not limited by the network capacity. This topology facilitates access for a large number of end users with a high key rate, utilizing a total fiber length of $L_1 + N \times L_2$. Assuming, without loss of generality, that $L_0 = L_1 + L_2$, our strategy can save a fiber length of $(N-1) \times L_1$ while achieving the same functionality.

The proposed protocol can also reduce fiber deployment in more complex scenarios. For instance, when four network participants aim to establish relay-less, end-to-end quantum-secured communication, a point-to-point strategy necessitates the deployment of at least six fiber links (see Fig. \ref{Fig: Discuss}(c)). However, by employing a broadcast strategy as depicted in Fig. \ref{Fig: Discuss}(d), Alice can simultaneously send quantum states to Charlie and David, and similarly, so can Bob. Consequently, link 1 and link 5 in Fig. \ref{Fig: Discuss}(c) can be eliminated without compromising network functionality. By appropriately positioning the beam splitter, the combined length of link 1 and link 2 in Fig. \ref{Fig: Discuss}(d) can be equivalent to the combined length of link 2 and link 4 in Fig. \ref{Fig: Discuss}(c), which significantly simplifies the network structure.

\onecolumngrid
\section*{Appendix}
\twocolumngrid


\section{Theoretical methods}
This section details the theoretical derivations. Based on the measurement model in Sec. \ref{Sec: 1A}, we prove the optimality of Gaussian attacks in the multi-user CV-QKD scenario (Sec. \ref{Sec: 1B}), resulting in a simplified key rate formula by security analysis with Gaussian states. We then provide the key rate calculation method using the covariance matrix (Sec. \ref{Sec: 1C}) and explain the independence and security of secret keys generated between Alice and the different Bobs in a 1-to-$N$ broadcast model (Sec. \ref{Sec: 1D}). In Sec. \ref{Sec: 1E}, we model the most practical broadcast channel built with beam splitters and analyze the most practical attack strategy, where Eve attacks each quantum channel independently. This yields an equivalent broadcast channel model, simplifying the analysis of attacks on the broadcast channel. Furthermore, we explain that the proposed protocol and security analysis method are independent of the channel model, enabling them to handle general attacks. Finally, we demonstrate through simulations how the proposed protocol minimizes the performance gap and show that the total key rate Alice can achieve with the multi-user CV-QKD scheme remains unaffected by the loss introduce by the broadcast operation (Sec. \ref{Sec: 1F}). This demonstrates the advantages of stable performance and high key rates, suggesting the potential for a CV-QKD network supporting a large number of users.

\subsection{Measurement model and entropy calculations}\label{Sec: 1A}
\begin{figure}
    \centering
    \includegraphics[width= 5 cm]{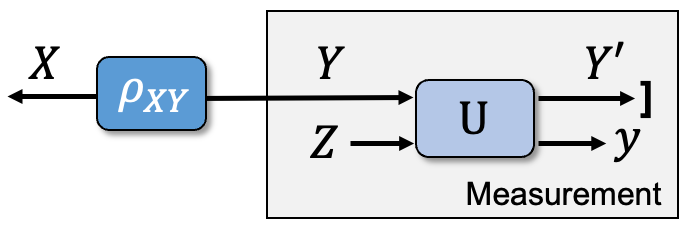}
    \caption{Measurement model. The mode $Y$ of the state $\rho_{XY}$ is measured. The measurement is modeled by a unitary operation with an ancillary system $Z$. The output modes include $Y'$ and $y$. $Y'$ is discarded and $y$ is preserved as the measurement result. 
    }\label{Fig: Measurement}
\end{figure}
The measurement operation on mode $Y$ in a bi-partite system $XY$ can be modeled by applying a unitary operation on $Y$ and an ancillary system $Z$ \cite{CVSecure2006second}, as shown in Fig. \ref{Fig: Measurement}. The output mode $y$ is the measurement results while $Y'$ is discarded. Without loss of generality, we consider that the ancilla has no correlation from the bi-partite system $XY$, i.e., $\rho_{XYZ} = \rho_{XY}\otimes \rho_{Z}$. Therefore, we have $S(XYZ)=S(XY)+S(Z)$ and $S(YZ)=S(Y)+S(Z)$. It indicates that 
\begin{equation}
    \begin{aligned}
        S(X:YZ) & = S(XYZ)-S(X)-S(YZ) \\
        & = S(XY) - S(X) - S(Y) \\
        & = S(X:Y).
    \end{aligned}
\end{equation}
Von Neumann entropy remains unchanged after a unitary operation, thus, we obtain $S(X:Y)=S(X:YZ)=S(X:Y' y)$. Based on the fact that discarding a system does not increase the Von Neumann entropy, we have 
\begin{equation}
    S(X:Y) = S(X:Y' y) \geq S(X:y).
\end{equation}
Since $S(X:Y) = S(X) - S(X|Y)$ and $S(X:y)=S(X)-S(X|y)$, we have
\begin{equation} \label{eq: conditional entropy}
    S(X|Y) \leq S(X|y).
\end{equation}

\subsection{Optimality of Gaussian attacks}\label{Sec: 1B}
Here we prove that Eve is optimized when the state of the network is a Gaussian state. We first detail the proof in a 3-party system, $ABC$, then extend it to a system with $N+1$ parties ($N>2$). 

\subsubsection{Proof in a 3-party system}
To prove that the Gaussian state $\rho_{ABC}^{G}$ optimizes Eve, it requires Eve's knowledge on Bob 1, $S(b : E c)$, to satisfy continuity, invariance under local Gaussification unitaries and strong subadditivity. In collective attacks, Eve can purify the system $\bar{A}\ \bar{B} \ \bar{C}$ with each mode extended to the mode with $M$ sub-modes (e.g., $\bar{A}=A^{1,2,...,M}$). Therefore, we consider the function $S(\bar{b}:E\bar{c})$ with respect to $\rho_{\bar{A}\ \bar{B} \ \bar{C}}$.

(i) Continuity: 
If $\| \rho^{(n)}_{\bar{A} \ \bar{B} \ \bar{C}} - \rho_{\bar{A} \ \bar{B}\ \bar{C}} \|_1 \leq \epsilon$, we can find a purification $\ket{\Psi}_{\bar{A} \ \bar{B} \ \bar{C} E}^{(n)}$ of $\rho^{(n)}_{\bar{A} \ \bar{B} \ \bar{C}}$ and a purification $\ket{\Psi}_{\bar{A} \ \bar{B} \ \bar{C} E}$ of $\rho_{\bar{A} \ \bar{B} \ \bar{C}}$, enabling $\| \hat{\Psi}_{\bar{A} \ \bar{B} \ \bar{C} E}^{(n)} - \hat{\Psi}_{\bar{A} \ \bar{B} \ \bar{C} E} \|_1 \leq 2 \sqrt{\epsilon}$ \cite{CVSecure2006second}. The partial trace can only decrease the trace norm indicates that $\| \rho^{(n)}_{\bar{b} \ \bar{c} E} - \rho_{\bar{b} \ \bar{c} E} \|_1 \leq 2 \sqrt{\epsilon}$, $\| \rho^{(n)}_{\bar{c} E} - \rho_{\bar{c} E} \|_1 \leq 2 \sqrt{\epsilon}$ and $\| \rho^{(n)}_{\bar{b}} - \rho_{\bar{b}} \|_1 \leq 2 \sqrt{\epsilon}$. Combining this result with the continuity of von Neumann entropies, we obtain the continuity of $S(\bar{b}:E\bar{c})$.

\begin{figure}
    \centering
    \includegraphics[width= 8.5 cm]{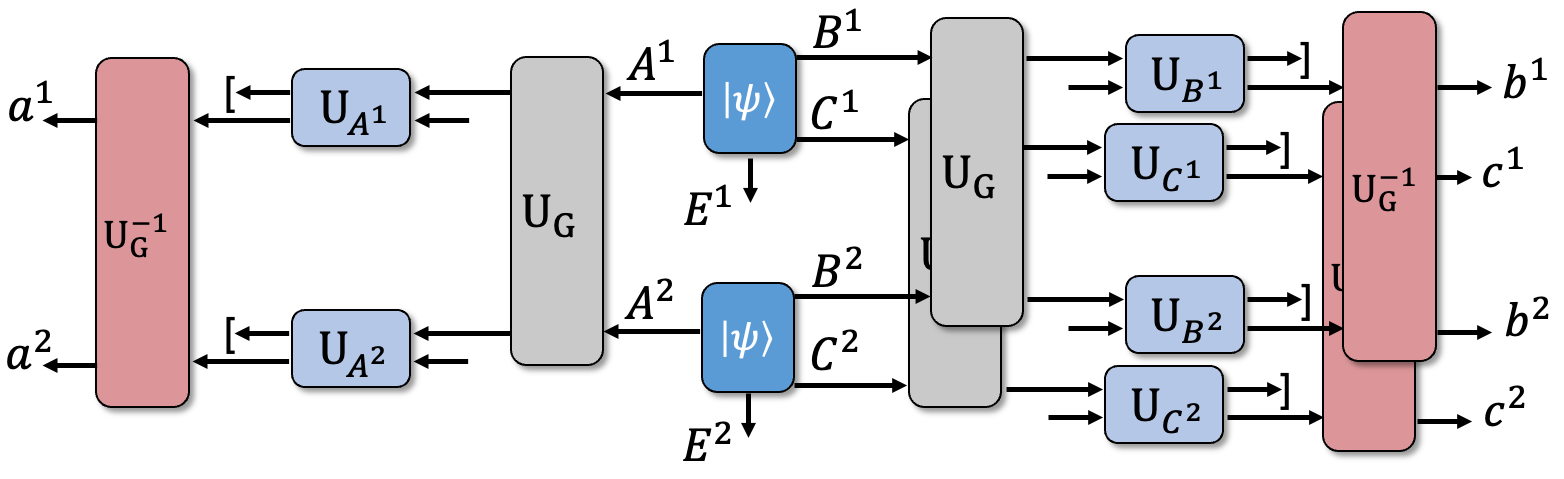}
    \caption{Invariance under local Gaussiﬁcation unitaries in a three-party system. Here, $M=2$. The measurement of a mode is modeled by a unitary operation with an ancilla and a partial trace. The Gaussiﬁcation operation can be interchanged with the unitary operation of the measurement, and be canceled by $U_{G}^{-1}$.
    }\label{Fig: LocalGauss}
\end{figure}

(ii) Invariance under local Gaussification unitaries: 
The local Gaussification operation for a two-party system, as described in \cite{CVSecure2006second}, can be expanded to a $3$-party system following the framework in \cite{wolf2006extremality}. Specifically, it is described by
\begin{equation}\label{eq: LocalGauss}
    {U^{\dagger}_{G}}\otimes{U^{\dagger}_{G}}\otimes{U^{\dagger}_{G}} \ \rho_{\bar{A} \ \bar{B}\ \bar{C}}^{\otimes M} U_G \otimes U_G \otimes U_G.
\end{equation}
Since Eve can purify each sub system $A^i B^i C^i E^i$, the Gaussification operation $U_G \otimes U_G \otimes U_G$ is equally applied to the global state $\ket{\psi}_{AB C E}^{\otimes M}$. After homodyne or heterodyne detection, the resulting state is $\widetilde{\rho} _{\bar{a}\bar{b} \bar{c} E}$. A simplified example for $M=2$ is illustrated in Fig. \ref{Fig: LocalGauss}, where $U_G$ is applied to the subsystems $A^1 A^2$, $B^1 B^2$ and $C^1 C^2$, respectively. Crucially, the unitary operation in a homodyne or heterodyne measurement can be interchanged with the Gaussification operation. This allows us to recover the product state $\rho_{abc}^{\otimes M}$ by applying ${U^{\dagger}_{G}}\otimes{U^{\dagger}_{G}}\otimes{U^{\dagger}_{G}}$. Notably, this product state can also be obtained by directly measuring $\ket{\psi}_{ABCE}^{\otimes M}$ without performing Gaussification operation. Since the mutual Von Neumann entropy is invariant under any local unitary operations, the desired invariance under local Gaussification unitaries for $S(\bar{b}:E\bar{c})$ is established.

(iii) Strong subaddivity:
We begin by proving the case for $M=2$. The generalization to $M>2$ follows straightforwardly. We consider the mutual entropy $S(b^{1}b^{2}:Ec^{1}c^{2})$, which can be expressed as 
\begin{equation}\label{eq:first decompos}
    S(b^{1}b^{2}:Ec^{1}c^{2}) = S(b^{1}b^{2}) - S(b^{1}b^{2}|Ec^{1}c^{2}).
\end{equation}
For the first item, the subadditivity of entropy implies $S(b^{1}b^{2}) \leq S(b^{1}) + S(b^{2})$. For the second item, we expand the conditional entropy as 
\begin{equation}\label{eq: second decompos}
    \begin{aligned}
        & S(b^{1}b^{2}|Ec^{1}c^{2}) \\
        & = S(b^{1}|b^{2}Ec^{1}c^{2}) + S(b^{2}|b^{1}Ec^{1}c^{2}) + S(b^{1}:b^{2}|Ec^{1}c^{2}) \\
        & \geq S(b^{1}|b^{2}Ec^{1}c^{2}) + S(b^{2}|b^{1}Ec^{1}c^{2}).
    \end{aligned}
\end{equation}
With Eq. \ref{eq: conditional entropy}, we have
\begin{equation}\label{eq: third decompos}
    \left\{
    \begin{aligned}
        & S(b^{1}|b^{2}Ec^{1}c^{2}) \geq S(b^{1}|B^{2}C^{2}Ec^{1}) \geq S(b^{1}|A^2 B^{2}C^{2}Ec^{1}) \\
        & S(b^{2}|b^{1}Ec^{1}c^{2}) \geq S(b^{2}|B^{1}C^{1}Ec^{2}) \geq S(b^{2}|A^1 B^{1}C^{1}Ec^{2})
    \end{aligned}
    \right. .
\end{equation}
Based on the security analysis in the main text, we have 
\begin{equation}
    \begin{aligned}
        & S(b^{1}:A^2 B^{2}C^{2}Ec^{1}) =  I(b^{1}:c^{1})\\
        & +\int dc^{1} \ p(c^{1}) S(\rho_{A^2 B^{2}C^{2}E}^{c^{1}}) \\ 
        & -\int db^{1} \ p(b^{1})\int dc^{1} \ p(c^{1}|b^{1})S(\rho_{A^2 B^{2}C^{2}E}^{b^{1}c^{1}}).
    \end{aligned}
\end{equation}
Since Eve can purify the system $\bar{A}\ \bar{B} \ \bar{C}$, the system $A^1 A^2 B^1 B^2 C^1 C^2 E$ is a pure state. When $b^{1}$ and $c^{1}$ are obtained with rank-1 measurements, the system $A^1 A^2 B^1 B^2 C^2 E$ (after the measurement of $C^1$), and the system $A^1 A^2 B^2 C^2 E$ (after the measurement of $B^1$ and $C^{1}$), are also pure states. Therefore, we have 
\begin{equation}
    \left\{
    \begin{aligned}
        & S(\rho_{A^2 B^{2}C^{2}E}^{c^{1}}) = S(\rho_{A^1 B^{1}}^{c^{1}}) \\
        & S(\rho_{A^2 B^{2}C^{2}E}^{b^{1}c^{1}}) = S(\rho_{A^1}^{b^{1}c^{1}})
    \end{aligned}
    \right. .
\end{equation}
Note that, the purification of $A^1 B^{1}C^{1}$ is $E^1$. Therefore, we have $S(\rho_{A^1 B^{1}}^{c^{1}}) = S(\rho_{E^1}^{c^{1}})$ and $S(\rho_{A^1}^{b^{1}c^{1}}) = S(\rho_{E^1}^{b^{1}c^{1}})$. It indicates that $S(b^{1}:A^2 B^{2}C^{2}Ec^{1}) = S(b^{1}:E^1 c^{1})$. Since $S(b^{1}:E^1 c^{1}) = S(b^{1}) - S(b^1 | E^1 c^{1})$ and $S(b^{1}:A^2 B^{2}C^{2}E c^{1}) = S(b^{1}) - S(b^1 | A^2 B^{2}C^{2}E c^{1})$, we obtain 
\begin{equation}\label{eq: subresult 1}
    S(b^1 | A^2 B^{2}C^{2}E c^{1}) = S(b^1 | E^1 c^{1}).
\end{equation}
Similarly, we have
\begin{equation}\label{eq: subresult 2}
    S(b^{2}|A^1 B^{1}C^{1}Ec^{2}) = S(b^{2}|E^2 c^{2}).
\end{equation}
Combining Eq. \ref{eq:first decompos}, Eq. \ref{eq: second decompos}, Eq. \ref{eq: third decompos}, Eq. \ref{eq: subresult 1} and Eq. \ref{eq: subresult 2}, we arrive at
\begin{equation}
    \begin{aligned}
        & S(b^{1}b^{2}:Ec^{1}c^{2}) \\
        & \leq S(b^{1}) + S(b^{2}) - S(b^{1}|E^1 c^{1}) - S(b^{2}|E^2 c^{2})\\
        & = S(b^{1} : E^1 c^{1}) + S(b^{2} : E^2 c^{2}).
    \end{aligned}
\end{equation}
This establishes the strong subadditivity property. $\square$

\subsubsection{Proof with arbitrary number of participants}
To extend the proof to a system with $N+1$ parties, we consider the system $A B_1 B_2 ... B_{N} E$. When analyzing Eve's knowledge about $b_1$, we denote the rest Bobs as $\mathbf{B_r} = B_2 B_3 ... B_{N}$ for simplicity. These $N-1$ modes are measured respectively, resulting in the measurement results $\mathbf{b_r}$. It is straightforward that $S(b_1 : E \mathbf{b_r})$ satisfies continuity as a function of state $\rho_{A B_1 B_2 B_3 ... B_{N}}$. To prove the invariance under local Gaussification unitaries, we can apply the same method as in the 3-party system. The only difference is that the local Gaussification operation is applied to $N+1$ modes, i.e., $A B_1 B_2 B_3 ... B_{N}$, with Eq. \ref{eq: LocalGauss} corrected to
\begin{equation}
    {U^{\dagger}_{G}}^{\otimes (N+1)} \ {\rho_{A B_1 B_2 B_3 ... B_{N}}}^{\otimes M} \ {U_G}^{\otimes (N+1)}.
\end{equation}
Here, the operation $U_G$ with $M$ input modes is applied to the $N$ modes ($A B_1 B_2 ... B_N$) respectively. This scheme aligns with the scheme in Ref. \cite{wolf2006extremality}, that the Gaussification operation respectively applied to different modes of the system. It still allows the interchange between a recovery operation $U_G ^{-1}$ that cancels the Gaussification operation and the measurement. Therefore, with the invariance of mutual Von Neumann entropy under local unitary operations, we have $S(b_1 : E \mathbf{b_r})$ is invariant under local Gaussification unitaries. Proving the strong subadditivity of $S(b_1 : E \mathbf{b_r})$ is straightforward by replacing the registers $C$ with $\mathbf{B_r}$ and $c$ with $\mathbf{b_r}$ in the above proofs. Here, we treat $\mathbf{B_r} = B_2 B_3 ... B_{N}$ as a whole: A `big' measurement is applied to $\rho_{\mathbf{B_r}}$, and the measurement result is $\mathbf{b_r}$.

\subsection{Key rate calculation based on covariance matrix}\label{Sec: 1C}
The key rate between Alice and Bob $i$ in the protocol can be estimated by 
\begin{equation}\label{eq: key rate}
    K_{AB_i} = \beta I(a:b_i) - I(b_i:\mathbf{b_r}) - S(b_i : E | \mathbf{b_r}).
\end{equation} 
Here, reconciliation efficiency is determined by error correction codes and the signal-to-noise ratio of the measurement results. The classical mutual information $I(a:b_i)$ and $I(b_i:\mathbf{b_r})$ can be achieved with the raw data experimentally achieved. Here, $\mathbf{b_r}$ represents the mode of the remaining users (e.g., $\mathbf{b_r}=b_{2,3,...,N}$ in a network with $N$ Bobs when $i=1$). $S(b_i : E | \mathbf{b_r})$ can be calculated with the covariance matrix of $\rho_{AB_i \mathbf{B_r}}$, which can be estimated with the raw data that can be experimentally collected. 

For Gaussian distributed raw data, we have 
\begin{equation}
    \begin{aligned}
        I(a:b_i) = 0.5 log_2 V_{B_{ix}} - 0.5 log_2 V_{B_{ix}|A_{x}} \\ + 0.5 log_2 V_{B_{ip}} - 0.5 log_2 V_{B_{ip}|A_{p}},
    \end{aligned}
\end{equation}
and
\begin{equation}
    \begin{aligned}
        I(b_i:\mathbf{b_r}) = 0.5 log_2 V_{B_{ix}} - 0.5 log_2 V_{B_{ix}|B_{rx}} \\ + 0.5 log_2 V_{B_{ip}} - 0.5 log_2 V_{B_{ip}|B_{rp}}.
    \end{aligned}
\end{equation}
Here, the conditional variance $V_{B_{ix}|B_{rx}}$ can be achieved from the covariance matrix $\gamma _{B_{ix}B_{rx}}$, namely, $V_{B_{ix}|B_{rx}} = V_{B_{ix}} - \sigma^T \gamma_{B_{rx}}^{-1} \sigma$. $\gamma_{B_{rx}}$ is the covariance matrix of the $x$-quadrature measurement results of the remaining Bobs, which forms $\gamma _{B_{ix}B_{rx}}$ with $\sigma$ as below,
\begin{equation}
    \gamma _{B_{ix}B_{rx}} = 
    \begin{bmatrix}
        V_{B_{ix}}  & \sigma^T \\
        \sigma    & \gamma_{B_{rx}}
    \end{bmatrix}.
\end{equation}
$V_{B_{ip}|B_{rp}}$ can be achieved in the same way.

For Gaussian state $\rho^G _{AB_i \mathbf{B_r}}$, using the fact that Eve can purify system $AB_i \mathbf{B_r}$, and $b_i , \mathbf{b_r}$ are the measurement results of modes $B_i , \mathbf{B_r}$ we have
\begin{equation}
    \begin{aligned}
        & S(b_i : E | \mathbf{b_r}) \\
        & = S(E|\mathbf{b_r}) - S(E|b_i \mathbf{b_r})\\
        & = S(AB_i|\mathbf{b_r}) - S(A|b_i \mathbf{b_r}) \\
        & = S(\rho_{AB_i}^{{\mathbf{b_r}}}) - S(\rho_{A}^{{b_i \mathbf{b_r}}}).
    \end{aligned}
\end{equation}
Here, $S(\rho_{AB_i}^{{\mathbf{b_r}}})$ and $S(\rho_{A}^{{b_i \mathbf{b_r}}})$ can be obtained from the symplectic eigenvalues of $\gamma_{AB_i}^{m_{\mathbf{B_r}}}$ and $\gamma_{A}^{m_{B_i \mathbf{B_r}}}$. These two matrixes can be obtained from the covariance matrix $\gamma_{AB_i \mathbf{B_r}}$ following the method in \cite{fossier2009improvement}.

Experimentally, we construct the covariance matrix $\gamma_{AB_i \mathbf{B_r}}$ with the $x$-quadrature and $p$-quadrature measurement results of modes $A, B_1 , B_2 , ..., B_{N} $. To model the impact of the detector imperfections, the limited detection efficiency and electronic noise can be calibrated in advance. This contribute to one \cite{zhang2020one} or two series \cite{fossier2009improvement} of trusted modes, representing the calibrated detector imperfections. Specifically, when one-time shot noise unit calibration is used \cite{zhang2020one}, we have $S(AB_i|\mathbf{b_r}) = S(\rho_{A B_i \mathbf{D}}^{\mathbf{b_r}})$ and $S(A|b_i \mathbf{b_r}) = S(\rho_{A \mathbf{D}}^{{b_i \mathbf{b_r}}})$. Here, mode $\mathbf{D}=D_1 D_2 ... D_N$ characterize the detection efficiencies that Eve cannot benefit from. 

\subsection{Independence and security of the keys}\label{Sec: 1D}
The practical significance of the proposed protocol lies in allowing Alice to simultaneously generate secret keys with all Bobs, requiring no restrictions on the users' operations or Eve's behavior. In this situation, when generating secret keys with Bob $1$, Eve is allowed to achieve the other Bobs' measurement results $\mathbf{b_r} = b_2 b_3 ... b_{N}$ and syndromes $\mathbf{s} = s(b_2) s(b_3) ... s(b_{N})$. Indeed, Eve and the other Bobs form a joint system, which can be expressed as $E \mathbf{b_r} \mathbf{s}$. The secret key generated between Alice and Bob $1$ remains secure against this joint system with Eq. \ref{eq: key rate}, ensuring the independence and security simultaneously. Since all Bobs' received states are transformed from the same state prepared by Alice, the mutual information $I(b_1:\mathbf{s})$ and $I(b_1:\mathbf{b_r})$ are non-negative. It indicates that holding the other Bobs' raw keys or syndromes could benefit Eve. 

Below we prove that the system $E \mathbf{b_r} \mathbf{s}$ holds the same amount of information about $b_1$ as $E \mathbf{b_r}$. Note that, $\mathbf{s}$ is the local function of $\mathbf{b_r}$, thereby we have $S(\mathbf{s} | \mathbf{b_r}) = 0$. Further, we have
\begin{equation}
    S(E \mathbf{b_r} \mathbf{s}) = S(E \mathbf{b_r}) +  S(\mathbf{s} | E \mathbf{b_r}) = S(E \mathbf{b_r}),
\end{equation}
and
\begin{equation}
    S(b_1 E \mathbf{b_r} \mathbf{s}) = S(b_1 E \mathbf{b_r}) + S(\mathbf{s} | b_1 E \mathbf{b_r}) = S(b_1 E \mathbf{b_r}).
\end{equation}
Therefore, we obtain
\begin{equation}
    \begin{aligned}
        S(b_1:E \mathbf{b_r} \mathbf{s}) & = S(b_1) + S(E \mathbf{b_r} \mathbf{s}) - S(b_1 E \mathbf{b_r} \mathbf{s})\\
        & = S(b_1) + S(E \mathbf{b_r}) - S(b_1 E \mathbf{b_r})\\
        & = S(b_1 : E \mathbf{b_r}).
    \end{aligned}
\end{equation}
It indicates that holding the syndromes $\mathbf{s}$ does not increase Eve's knowledge about $b_1$ when Eve is assumed to have the other Bobs' measurement results $\mathbf{b_r}$. 

$S(b_1 : E \mathbf{b_r})$ characterizes the accessible information about $b_1$ when Eve and the rest Bobs are in collaboration. Therefore, the secret key rate $K_{AB_1} = \beta I(a:b_1) - S(b_1 : E \mathbf{b_r})$ ensures that both Eve and the other Bobs are unable to hold knowledge about the secret key generated between Alice and Bob 1. This offers the security and independence simultaneously. Moreover, it indicates that, Alice can make full use of the entire system $AB_1 B_2 ... B_{N}$ and achieve a total key rate of $K_A = \sum_{i=1}^N K_{AB_i}$.

\begin{figure}[t]
    \centering
    \includegraphics[width= 6.5 cm]{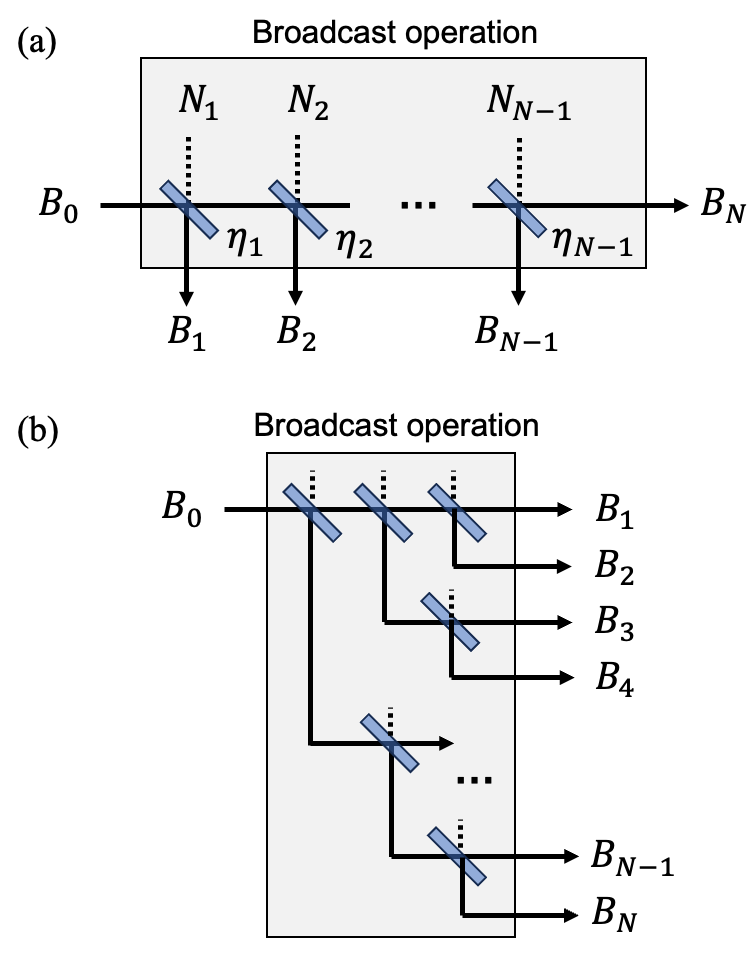}
    \caption{Broadcast operation model with beam splitters. (a) A model with $N-1$ beam splitters for realizing a 1-to-$N$ broadcast. (b) An equivalent model with tree-type connected beam splitters. Here, $N_i$ is the mode of vacuum state.}\label{Fig: Broadcast}
\end{figure}

\subsection{Broadcast channel model with beam splitters}\label{Sec: 1E}
The broadcast channel required in the protocol can be modeled with several beam splitters, enabling the mode prepared by Alice, $B_0$, to be split into $N$ modes, $B_1 B_2 ... B_{N}$. 

\subsubsection{Basic model of broadcast operation}

The model of the broadcast operation is illustrated in Fig. \ref{Fig: Broadcast}. To broadcast the mode $B_0$ to $N$ Bobs, $N-1$ beam splitters are used, as shown in Fig. \ref{Fig: Broadcast} (a). Each beam splitter has a transmittance of $\eta_i$. The transmittance between $B_0$ and arbitrary mode $B_i$ is $\Pi_{j=1}^{i-1} \eta_j (1-\eta_i)$. To realize a uniform broadcast operation, i.e. the transmittance between $B_0$ and arbitrary $B_i$ is $1/N$, it requires $\eta_i = 1 - 1/(N-i+1)$. It is equivalent to a tree-type connected beam splitter model, as shown in Fig. \ref{Fig: Broadcast} (b). 

After the $1-to-N$ uniform broadcast operation, the variance of the output mode $B_i$ is
\begin{equation}
    V_{B_i} = \frac{1}{N} V_{B_0} + 1 - \frac{1}{N}.    
\end{equation}
The covariance between any two output modes, e.g., $B_i$ and $B_j$ is 
\begin{equation}
    C_{B_i B_j} = \frac{1}{N} (V_{B_0} - 1).
\end{equation}
The covariance between the output mode $B_i$ and any other mode, e.g., $A$ is
\begin{equation}
    C_{A B_i} = \sqrt{\frac{1}{N}} C_{AB_0}.
\end{equation}
Here, $V_{B_0}$ is the variance of mode $B_0$, and $C_{AB_0}$ is the covariance between mode $A$ and mode $B_0$.

\subsubsection{Channel-by-channel attack and simplified broadcast channel model}

\begin{figure}[t]
    \centering
    \includegraphics[width= 6.5 cm]{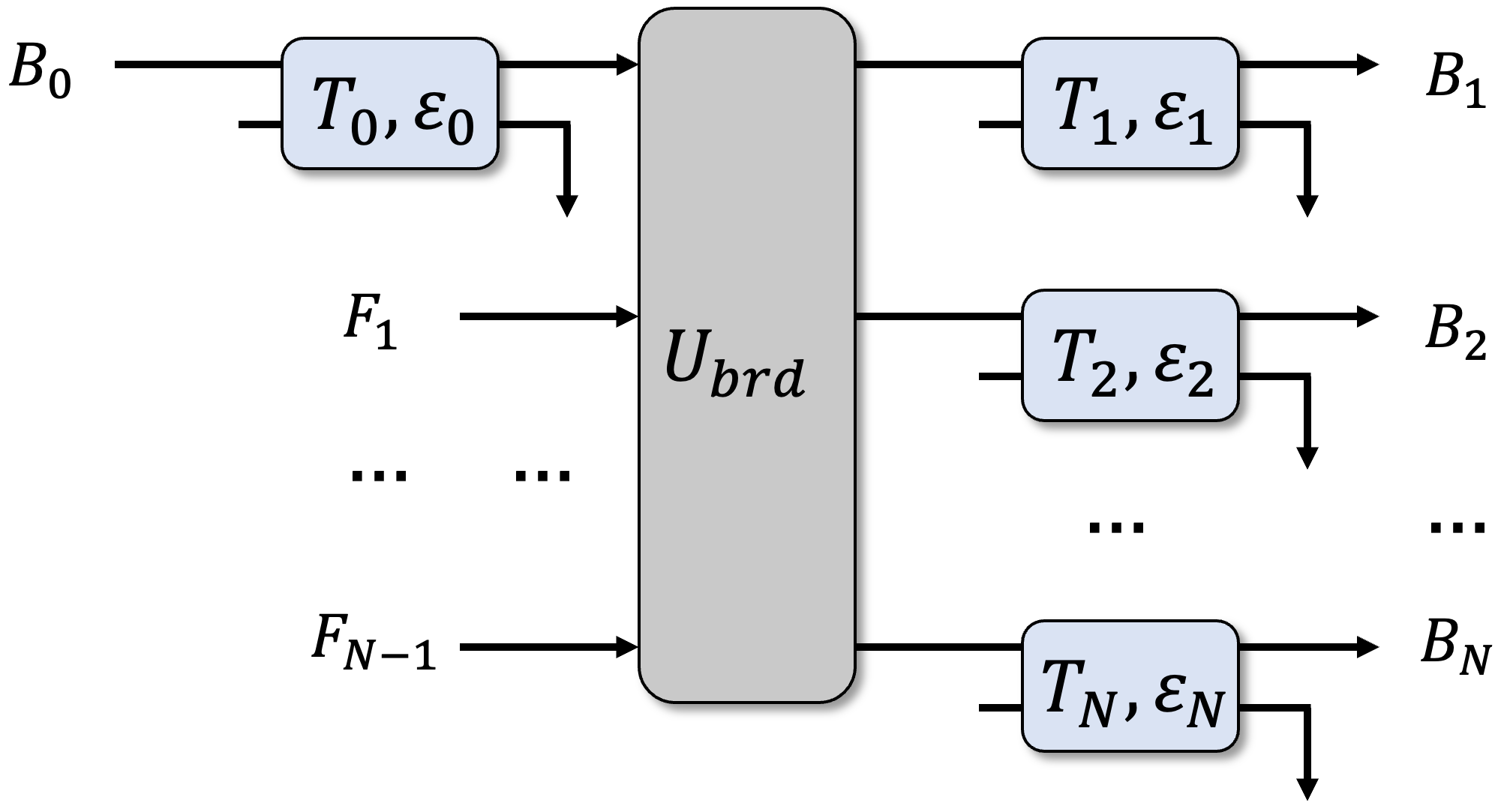}
    \caption{Channel-by-channel attack model. $U_{brd}$ is the broadcast operation, and $F_i$ is the ancillary mode. $T_i$ represents the channel transmittance, and $\varepsilon_i$ represents the excess noise. Eve attacks the channel between $B_0$ and $U_{brd}$, and the channels between $U_{brd}$ and $B_1 B_2 ... B_N$ respectively.}\label{Fig: ChbyChAttack}
\end{figure}

As depicted in Fig. \ref{Fig: ChbyChAttack}, the most direct attack Eve can implement is a channel-by-channel attack. In this scenario, the initial mode $B_0$ passes a channel under Eve's control, characterized by a transmittance $T_0$ and excess noise $\varepsilon_0$. Subsequently, a broadcast operation is applied, resulting in $N$ output modes. Each of these modes then propagates through individual channels manipulated by Eve, each with a transmittance $T_i$ and excess noise $\varepsilon_i$. The resulting output modes are $B_1, B_2, ..., B_N$, which are ultimately measured by distinct Bobs. Eve independently targets the channel connecting $B_0$ and $U_{brd}$, as well as each of the channels between $U_{brd}$ and the respective output modes $B_1, B_2, ..., B_N$.

\begin{figure}
    \centering
    \includegraphics[width= 8.5 cm]{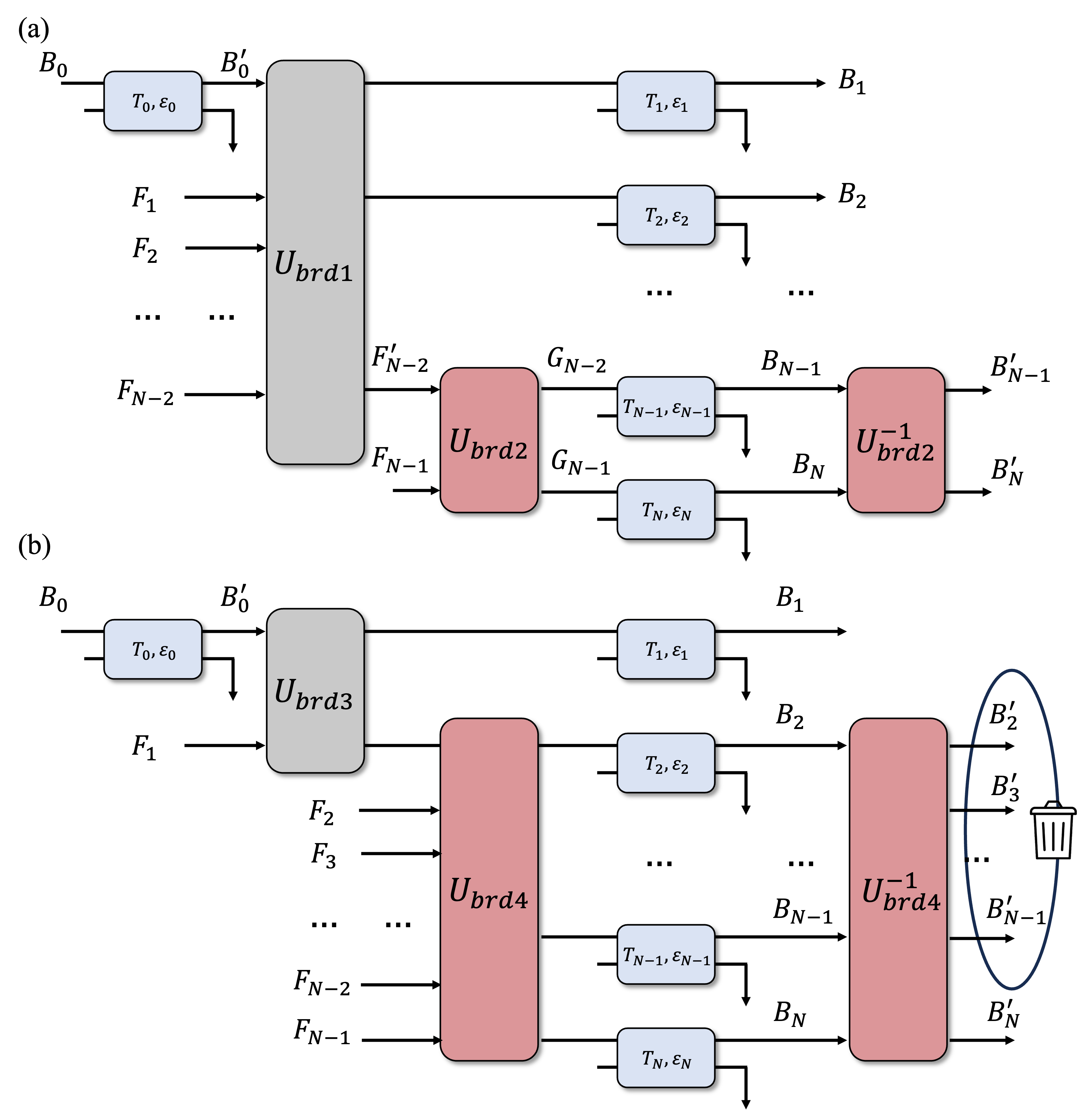}
    \caption{Simplification of the channel-by-channel attack model. (a) $U_{brd}$ consists of $U_{brd1}$ and $U_{brd2}$. A reverse operation $U_{brd2}^{-1}$ canceled $U_{brd2}$, ensuring that $B_{N-1}'$ has no correlation with the other Bobs' modes. (b) $U_{brd}$ consists of $U_{brd3}$ and $U_{brd4}$. A reverse operation $U_{brd4}^{-1}$ canceled $U_{brd4}$, ensuring that $B_2 ', B_3 ' ... B_{N-1}'$ has no correlation with $B_N '$. $T_i$ represents the channel transmittance, and $\varepsilon_i$ represents the excess noise. }\label{Fig: Simplification}
\end{figure}

We can further simplifies the broadcast channel model by applying a reverse unitary operation to any two Bobs' modes, e.g., $B_{N-1}$ and $B_{N}$, to cancel a part of the broadcast operation (see Fig. \ref{Fig: Simplification} (a)). Specifically, the 1-to-$N$ broadcast operation can be modeled by two sub-operations, $U_{brd1}$ and $U_{brd2}$. Here, $U_{brd1}$ broadcasts one mode to $N-1$ modes, and $U_{brd2}$ broadcasts one mode to two modes. $U_{brd2}^{-1}$ cancels $U_{brd2}$ by making $B_{N}'$ uncorrelated with all other Bobs' modes, therefore we can discard mode $B_{N}'$ while remaining the same secret key rate. 

The simplest $U_{brd2}$ is a beam splitter with transmittance $\eta$, which results in 
\begin{equation}
    G_{N-2} = \sqrt{\eta}F_{N-2}' + \sqrt{1-\eta}F_{N-1},
\end{equation}
and 
\begin{equation}
    G_{N-1} = \sqrt{\eta}F_{N-1} - \sqrt{1-\eta}F_{N-2}'.
\end{equation}
We use a beam splitter with transmittance of
\begin{equation} \label{eq:eta}
    \eta'= (1-\eta)T_{N} / (\eta T_{N-1} + (1-\eta) T_{N})
\end{equation}
to realize the reverse operation $U_{brd2}^{-1}$. It decouples $B_{N-1} '$, resulting in a zero covariance between $B_{N-1} '$ and any other modes, e.g., $A$. Specifically,
\begin{equation}\label{eq:C}
    \begin{aligned}
        C_{AB_{N-1} '} & = \sqrt{\eta'}C_{AB_{N-1}} + \sqrt{1-\eta'}C_{AB_{N}} \\
        & = \sqrt{\eta' T_{N-1}}C_{AG_{N-2}} + \sqrt{(1-\eta') T_{N}}C_{AG_{N-1}}.
    \end{aligned}
\end{equation}
Here, we have $C_{AG_{N-2}} = \sqrt{\eta}C_{AF_{N-2}'}$ and $C_{AG_{N-1}} = -\sqrt{1-\eta}C_{AF_{N-2}'}$, since $C_{AF_{N-1}} = 0$. Combining with Eq. \ref{eq:eta} and Eq. \ref{eq:C}, we have
\begin{equation}
    \begin{aligned}
        & C_{AB_{N-1} '} \\
        & = (\sqrt{\eta' \eta T_{N-1}} - \sqrt{(1-\eta') (1-\eta) T_{N}}) C_{AF_{N-2}'} \\
        & = 0.
    \end{aligned}
\end{equation}
It indicates that $\rho_{AB_1 B_2 ... B_{N-1}' B_{N}'} = \rho_{AB_1B_2 ... B_{N}'} \otimes \rho_{B_{N-1}'}$. When calculating the secret key rate between Alice and Bob 1, since the local unitary operation does not change the mutual information, we have 
\begin{equation}
    I(b_1 : b_2 b_3 ... b_{N-1} b_{N}) = I(b_1 : b_2 b_3 ... b_{N-1}' b_{N}'),
\end{equation}
and
\begin{equation}
    S(b_1 : E | b_2 b_3 ... b_{N-1} b_{N}) = S(b_1 : E | b_2 b_3 ... b_{N-1}' b_{N}').
\end{equation}
Further, since mode $b_{N-1}'$ is independent to the network system, we have 
\begin{equation}
    I(b_1 : b_2 b_3 ... b_{N-2} b_{N-1}' b_{N}') = I(b_1 : b_2 b_3 ... b_{N-2} b_{N}'),
\end{equation}
and
\begin{equation}
    S(b_1 : E | b_2 b_3 ... b_{N-2} b_{N-1}' b_{N}') = S(b_1 : E | b_2 b_3 ... b_{N-2} b_{N}').
\end{equation}
Therefore, we can obtain an equivalent channel between mode $F_{N-2}'$ and $B_{N-1}'$. The equivalent transmittance is 
\begin{equation}
    T_{eq} = \eta T_{N-1} + (1-\eta) T_{N},
\end{equation}
and the equivalent excess noise is
\begin{equation}
    \varepsilon_{eq} = \frac{\eta T_{N-1} ^2 \varepsilon_{N-1} + (1-\eta)T_N ^2 \varepsilon_{N}}{T_{eq}^2}.
\end{equation}
If $T_{N-1} = T_N = T'$, we have
\begin{equation}
    T_{eq} = T', \varepsilon_{eq} = \eta  \varepsilon_{N-1} + (1-\eta) \varepsilon_{N}.
\end{equation}
If $T_{N-1} = T_N = T'$, and $\varepsilon_{N-1} = \varepsilon_{N} = \varepsilon'$, we have
\begin{equation}
    T_{eq} = T', \varepsilon_{eq} = \varepsilon'.
\end{equation}
Based on this simplification strategy, we can then remove modes $B_{N-2}$, $B_{N-3}$, ..., $B_2$ by the similar reverse operations. The final channel model after the simplification is shown in Fig. \ref{Fig: Simplification} (b). Here, the 1-to-$N$ broadcast operation consists of $U_{brd3}$ and $U_{brd4}$, where $U_{brd3}$ realizes the 1-to-2 broadcast, while $U_{brd4}$ realizes the 1-to-$(N-1)$ broadcast. The modes $B_2$, $B_3$, ..., $B_{N-2}$, $B_{N-1}$ can be removed after the reverse operation $U_{brd4}^{-1}$. Finally, we only need to consider an equivalent 1-to-2 system, which significantly simplifies the protocol performance analysis under such channel-by-channel attacks.

\begin{figure}
    \centering
    \includegraphics[width= 8.5 cm]{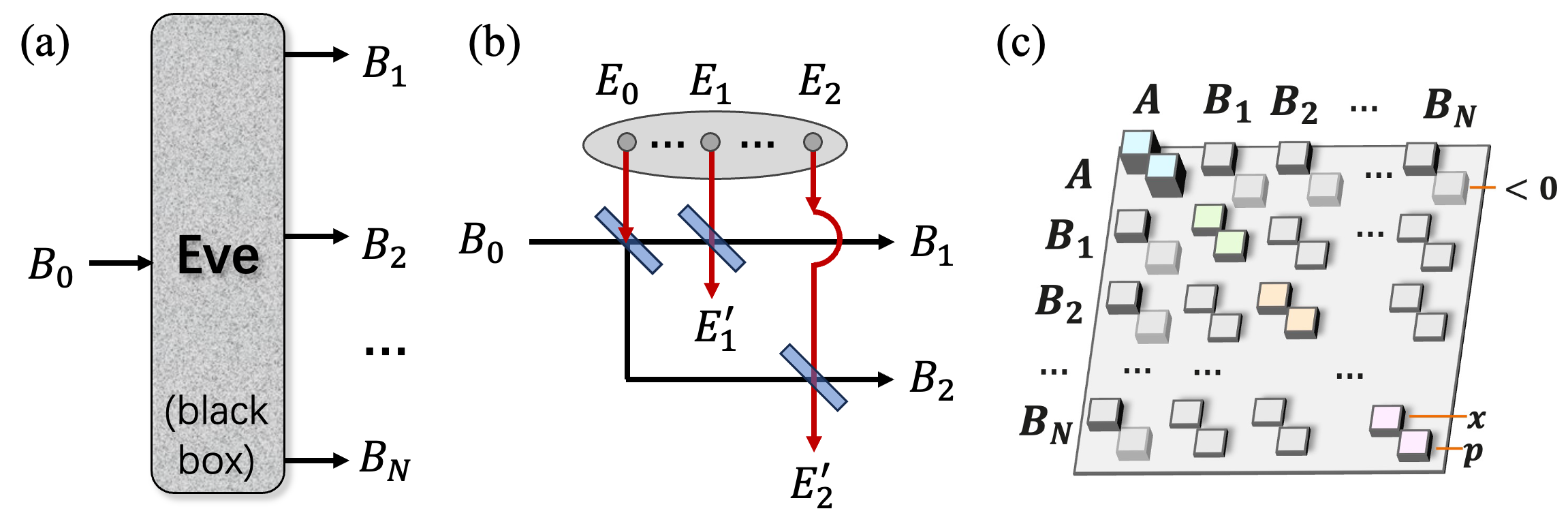}
    \caption{The general attack model and the covariance matrix. (a) The broadcast channel under a general attack. The broadcast channel is a black box for the legitimate parties in the network, and Eve fully controls the broadcast operation. (b) A typical attack introducing additional correlations between the two channels ($B_0 - B_1$ and $B_0 - B_2$) (c) The covariance matrix $\gamma _{AB_1 B_2 ... B_N}$ characterizing the overall multi-user system. }\label{Fig: GeneralAttack}
\end{figure}

\subsubsection{Security against general attacks}
Fig. \ref{Fig: GeneralAttack} (a) shows a general attack scheme. Eve fully controls the channel, while the legitimate parties (Alice and Bobs) in the multi-user system have no knowledge about Eve's operation. Eve can introduce additional correlations between different channels, as well as change the broadcast operation, to benefit herself. A typical attack strategy for a 1-to-2 broadcast channel is shown in Fig. \ref{Fig: GeneralAttack} (b), where Eve interacts mode $E_0$ with the first beam splitter (the broadcast operation), and implements a two-mode attack \cite{CVMDIYork} to the two channels by interacting modes $(E_1 , E_2)$ with the two channels. Modes $(E_0 , E_1 , E_2)$ are chosen from a set. This is a more sophisticated attack comparing with the aforementioned channel-to-channel attack and can provide Eve more information about the multi-user CV-QKD system.

The proposed security analysis method can handle the general attacks because it relies on the covariance matrix of the multi-user system, as shown in Fig. \ref{Fig: GeneralAttack} (c), not the channel model. As long as Bobs can achieve the measurement results and proceed with post-processing with Alice, they can construct the covariance matrix $\gamma_{AB_1 B_2 ... B_N}$ and achieves the secret key rates that satisfies independence and security simultaneously. Any operation by Eve will be reflected in the covariance matrix, allowing Alice and the Bobs to mitigate Eve's advantage and ensure security against general attacks.

\begin{figure*}
    \centering
    \includegraphics[width= 12.5 cm]{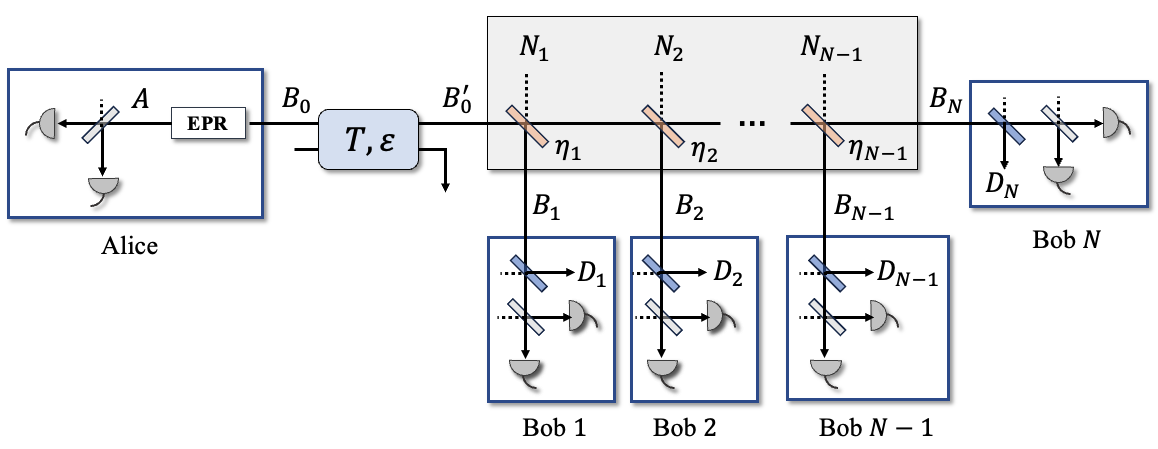}
    \caption{The entanglement-based scheme used in simulations. Alice holds an EPR state, with mode $A$ preserved and heterodyne detected, and mode $B_0$ sent to other Bobs through a broadcast channel. The equivalent channel parameters are transmittance $T$ and excess noise $\varepsilon$. Following the scheme introduced in Fig. \ref{Fig: Broadcast}, we use $N-1$ beam splitters with transmittance of $\eta_{i} =  1 - 1/(N-i+1)$ to realize the broadcast operation, transforming the input mode $B_0 '$ to $N$ output modes ($B_1 , B_2 , ... , B_N$). Each mode ($B_j$) received by a particular Bob is heterodyne detected respectively. The imperfections of the detection, including the limited detection efficiency ($\eta_d$) and electronic noise ($\nu_{ele}$), is modeled by a trusted beam splitter, with transmittance $\eta_{D}=\eta_{d}/(1+\nu_{ele})$ \cite{zhang2020one}. Without loss of generality, we assume that all Bobs' detector imperfections are the same. }\label{Fig: SimulMethod}
\end{figure*}

\subsection{Simulations}\label{Sec: 1F}
Here, we detail the simulation method, analyze the factors contributing to the performance gap between the proposed protocol and the achievable upper limit, and present the protocol performance across different scenarios.

\subsubsection{Performance gap with the upper limit}

Consistent with the conclusions in the main text, the achievable upper limit of the key rate of the 1-to-N CV-QKD is achieved when the $N$ Bobs locate in the same site. Equivalently, this upper limit corresponds to the key rate of a point-to-point protocol with channel parameters $(T, \varepsilon)$ equivalent to the multi-user scenario. Note that, with a proper reverse operation to cancel the broadcast operation, we can always obtain the equivalent channel parameters when the broadcast operation is unitary. In our simulations, we focus on the scenario with a broadcast operation consists of several beam splitters (discussed in Fig. \ref{Fig: Broadcast}), which aligns with practical implementations. The entanglement-based scheme we used for simulations is shown in Fig. \ref{Fig: SimulMethod}. One mode ($B_0$) of the ideal EPR state prepared by Alice is affected by the channel with transmittance $T$ and excess noise $\varepsilon$. Then, the output mode of the channel, $B_0 '$, is interacted with the beam-splitter-based broadcast operation, resulting in $N$ output modes. The legitimate parties' modes, including $A$, $B_1$, $B_2$, ..., $B_N$ are heterodyne detected respectively. The imperfect of each Bob's heterodyne detection is modeled by a beam splitter with transmittance $\eta_{D}=\eta_{d}/(1+\nu_{ele})$ \cite{zhang2020one}, which provides a series of additional modes for parameter estimation ($D_i$). Here, $\eta_d$ is the detection efficiency, and $\nu_{ele}$ is the electronic noise. Without loss of generality, all Bobs' detector imperfections are assumed to be the same.

\begin{figure} [b]
    \centering
    \includegraphics[width= 8.5 cm]{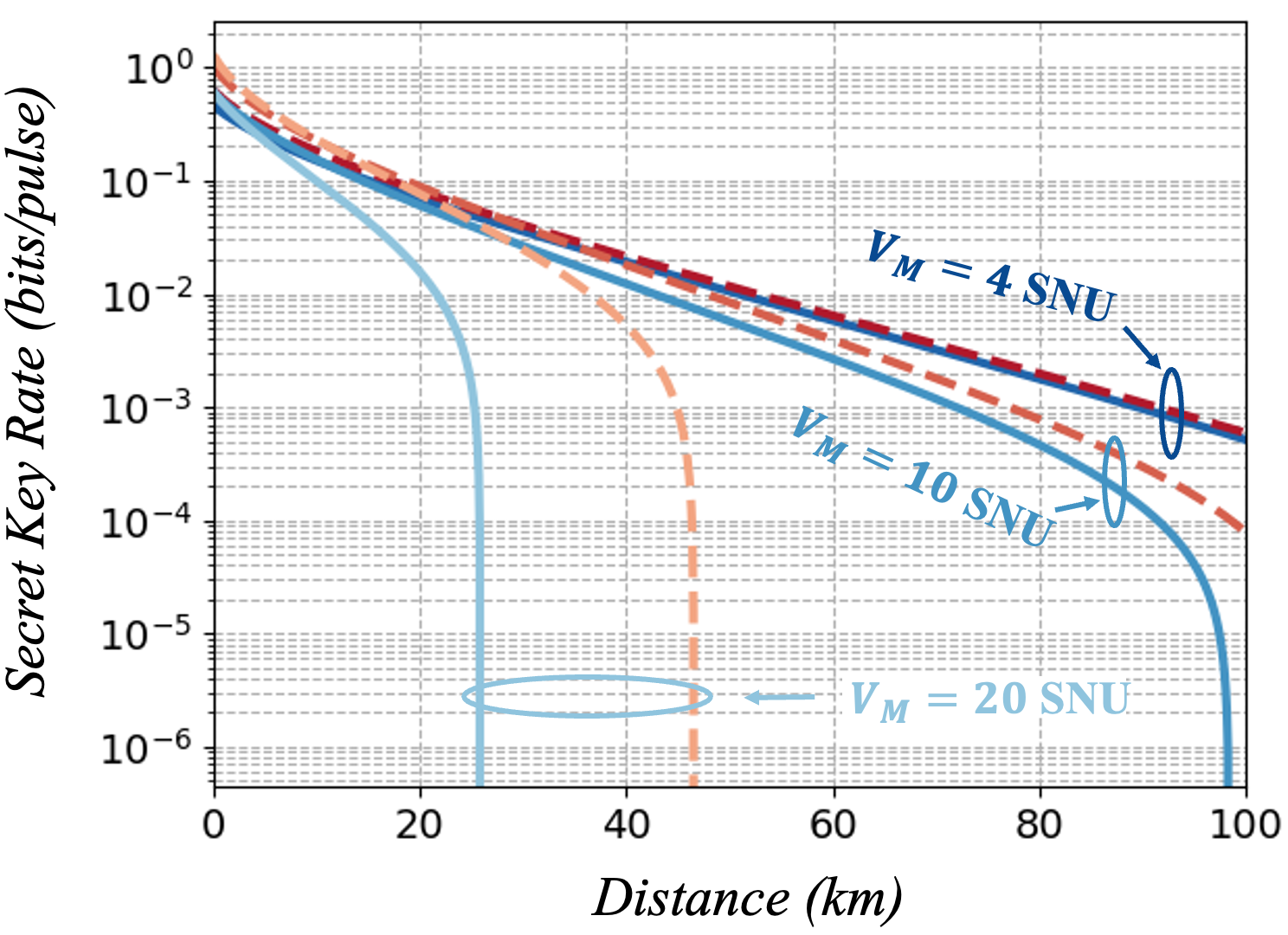}
    \caption{Protocol performance compared to upper limit with different modulation variance. Solid line: key rate of the proposed protocol, dashed line: the achievable upper limit in multi-user CV-QKD. Simulation parameters: $N = 4$, $\beta=0.956$, $\varepsilon=0.05$ SNU, detection efficiency $\eta=0.6$, electronic noise $\nu_{ele}=0.1$ SNU, $V_M = 4 / 10 / 20$ SNU.}\label{Fig: Simulation1}
\end{figure}

In ideal scenarios, a reconciliation efficiency of 1 is achievable, enabling an infinite optimal modulation variance. Therefore, we simulate the protocol performance in such case with a modulation variance of $10^4$ shot noise unit (SNU) and a pure loss channel ($\varepsilon=0$). To align with the optical fiber channel, we set the link loss at $\alpha=0.2$ dB/km. In practical situations, the reconciliation efficiency is normally 0.956 \cite{CvExp202kmPRL}, resulting in an optimal modulation variance of 4 SNU. It indicates that the average photon number of the quantum signals is 2, resulting in a significant low signal-to-noise ratio (SNR) at Bobs' sites, which contributes to the easy de-correlation. The proposed protocol is easier to reach the achievable upper limit in practical cases as shown in the main text, and the imperfections of the detectors also reduce the SNR and narrow the performance gap with the upper limit. 
As shown in Fig. \ref{Fig: Simulation1}, in practical situations, a higher modulation variance widens the key rate gap to the upper limit. For $V_M = 4$ SNU, the proposed protocol approaches the upper limit at all distances within 100 km, while for $V_M =$ 10 and 20 SNU, an obvious performance gap is introduced, and the maximum transmission distance is notably limited.

\subsubsection{Multi-user protocol performance in ideal conditions}
\begin{figure*}
    \centering
    \includegraphics[width= 17 cm]{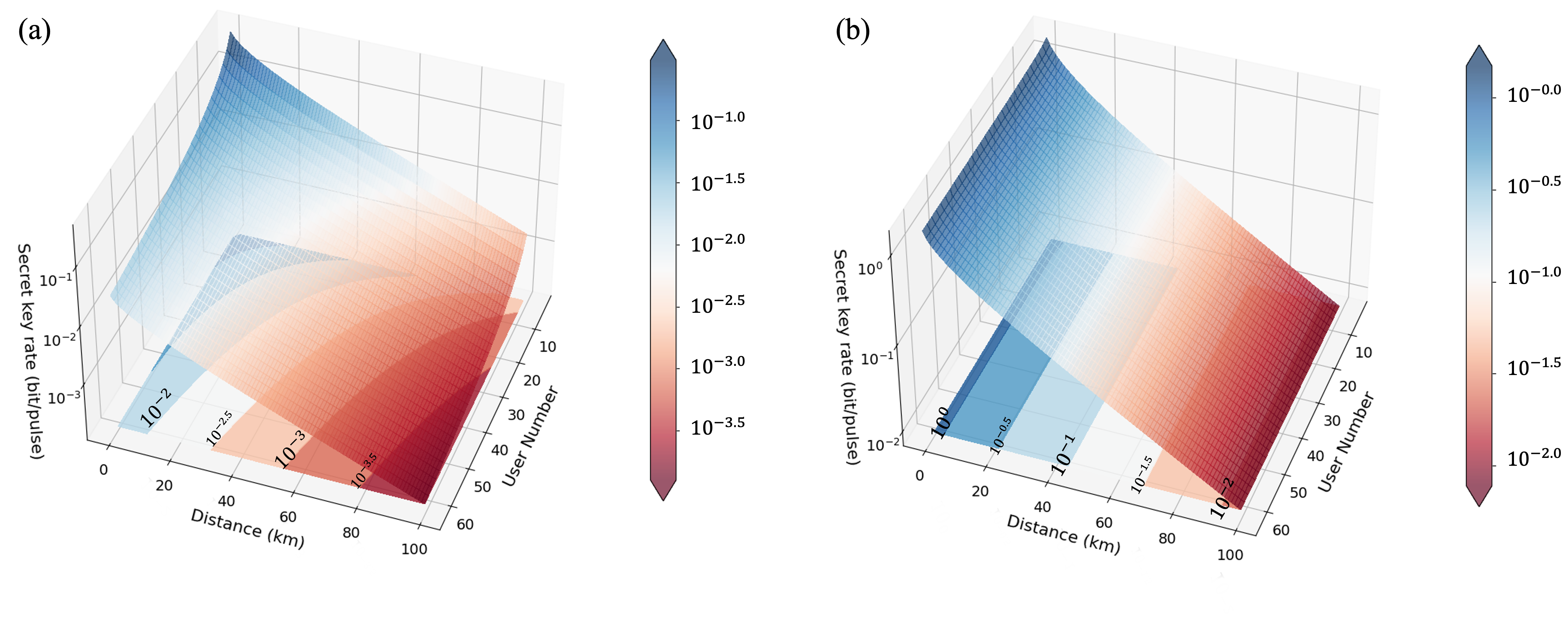}
    \caption{Visualization of the per-user key rate (a) and total key rate (b) in ideal situation. The 3D surface illustrates the secret key rate as a function of transmission distance and the number of users. The accompanying 2D projection displays the corresponding range of secret key rates across varying distances and user numbers. Simulation parameters: $\beta=1$, $V_M = 10^4$ SNU, $\varepsilon=0.0$ SNU, ideal detector.}\label{Fig: Simulation2}
\end{figure*}

Here, we present further simulation results for the proposed protocol under ideal conditions. Fig. \ref{Fig: Simulation2} (a) and (b) illustrate the per-user and total key rates in the ideal scenario. The per-user key rate exceeds $10^{-2.5}$ bit/pulse for access network distances (up to 25 km) and remains above $10^{-4}$ bit/pulse for metropolitan network transmission (up to 100 km). Notably, the total key rate remains stable despite an increasing number of users and the associated higher channel loss. For instance, while the broadcast loss increases from 6 dB for 4 users to 18 dB for 64 users in a 1-to-$N$ CV-QKD system, the proposed protocol's performance remains unaffected.

\subsection{Networks with physical user isolation}
Building on the framework presented in the main text, we now analyze networks with physical user isolation. Consider a network comprising a server, Alice, and multiple end users, Bob $i$, where $i = 1, 2, ..., N$. Physical user isolation implies that the state characterizing the entire network can be expressed as $\rho_{AB_1 B_2 ... B_N} = \otimes_i ^N \rho_{A_i B_i}$. Consequently, its end-to-end key rate follows a straightforward point-to-point formulation. Specifically, for the link between Alice and Bob 1, we have $K_{AB_1} = \beta I(a_1:b_1) - \chi(b_1 : E b_2 b_3 ... b_N) = \beta I(a_1:b_1) - \chi(b_1 : E)$. 

Indeed, many existing QKD network implementations align with this classification:
(1) Independent QKD links: In a network where Alice establishes separate links with different Bobs, each Alice-Bob $i$ link is characterized by the state $\rho_{A_i B_i}$. Collectively, $N$ such links form the overall network, expressed as $\rho_{AB_1 B_2 ... B_N} = \otimes_i ^N \rho_{A_i B_i}$.
(2) Frequency-division multiplexing: In a QKD network utilizing different frequency bands, independent QKD links are simultaneously established. Transmission in frequency band $i$ results in the state $\rho_{A_i B_i}$ between Alice and Bob $i$.
(3) Time-division multiplexing: In a QKD network employing time-division multiplexing, point-to-point QKD links are created in separate time slots, each establishing the state $\rho_{A_i B_i}$. Notably, single-photon QKD networks using a broadcast channel naturally achieve time-division multiplexing, yielding a decomposable multipartite system that simplifies analysis. In contrast, CV-QKD networks broadcasting a coherent state source produce a coupled system, as multi-photon signals can simultaneously activate all receivers at the end users' sites.


\section{Experimental methods}
\begin{figure*}
    \centering
    \includegraphics[width= 15 cm]{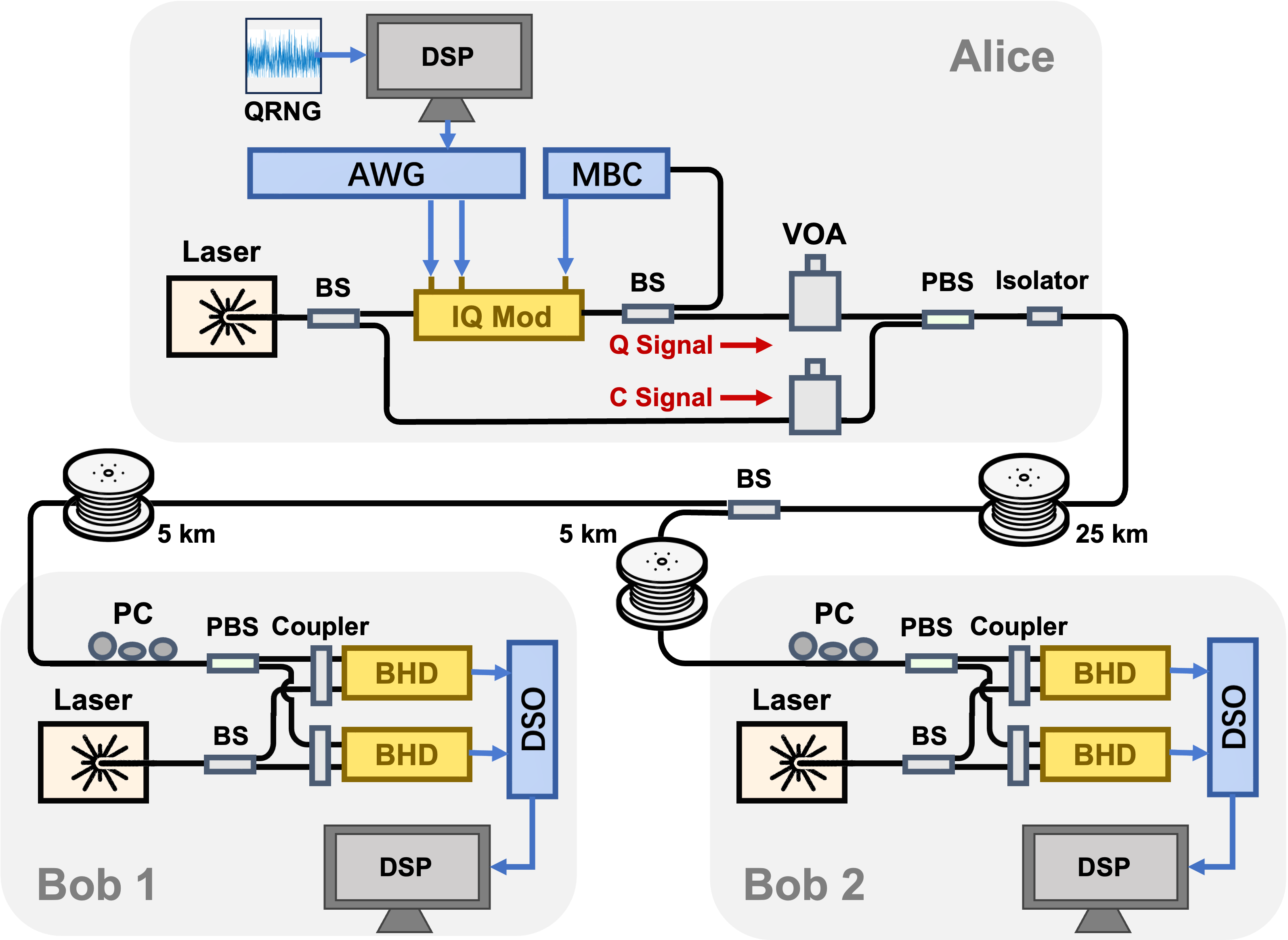}
    \caption{The setup of the multi-user CV-QKD experiment. Alice prepares Gaussian modulated coherent states and sends them to Bob 1 and Bob 2 through a broadcast channel. AWG: arbitrary waveform generator, MBC: modulator bias controller, DSO: digital storage oscilloscope, DSP: digital signal processing, BS: bean splitter, IQ Mod: In-phase/quadrature modulator, VOA: variable optical attenuator, PBS: polarization beam splitter, PC: polarization controller.}\label{Fig: ExpScheme}
\end{figure*}

\subsection{Optical setup and data processing}
Fig. \ref{Fig: ExpScheme} illustrates a 3-node metropolitan CV-QKD network in a 1-to-2 configuration, employing the proposed multi-user CV-QKD protocol. Alice utilizes a continuous-wave laser (NKT Photonics Basik X15) at 1550.12 nm with a 0.1 kHz linewidth as the optical carrier. The laser output is split by a beam splitter. One branch undergoes Gaussian modulation via an in-phase/quadrature (IQ) modulator (Fujitsu FTM7962EP) driven by an arbitrary waveform generator at a 30 GSa/s sampling rate. The modulation data is generated from the random numbers achieved from a quantum random number generator. After digital signal processing (DSP), the pre-processed data is sent to the arbitrary waveform generator. The IQ modulator has a 23 GHz bandwidth, and the system baud rate is 1 GBaud. A portion of the modulated signal is directed to a modulator bias controller, which keeps the modulator operated in a carrier suppression mode. Subsequently, a variable optical attenuator (EXFO FTBx-3500-BI) further reduces the quantum signal (Q signal) power to achieve an average photon number of 2. The variable optical attenuator incorporates a real-time optical power monitor module, enabling precise control of the average photon number. The remaining branch serves as the classical pilot signal (C signal), providing a phase reference for the receivers. The Q and C signals are then combined by a polarization beam splitter, positioned in orthogonal polarization directions and co-transmitted in the broadcast quantum channel.

The 1-to-2 broadcast channel comprises a 25 km optical fiber (SMF-28) followed by a beam splitter, with each output connected to a 5 km optical fiber (SMF-28). Alice's signal is split by the broadcast channel, enabling responses from both Bobs.
Each Bob first adjusts the polarization of the incoming signal by a polarization controller, then decouples the Q and C signals with a polarization beam splitter. Subsequently, the Q and C signals are detected by separate balanced homodyne detectors. It employs a real local oscillator (LO) scheme where each Bob uses an independent laser (NKT Photonics Basik X15) with 12 dBm optical power as the LO. Each LO laser's center frequency is offset by approximately 1.55 GHz from Alice's laser to enable intermediate-frequency coherent detection. The $x$ and $p$ quadratures of the Q signal are retrieved from the detection results of one homodyne detector, while the other provides the frequency difference between the LO and the optical carrier for phase recovery. A digital storage oscilloscope at each Bob's site collects the detection results from both homodyne detectors. Following DSP at Bob's site, noise-suppressed measurement results are obtained.

DSP is used to avoid the base band noise and to suppress the excess noise caused by inter-symbol interference, fast-fading phase difference between the LO and optical carrier, and the slow-fading phase noise. Initially, Alice upconverts the baseband signal (centered at 0 Hz) to an intermediate frequency band (centered at 750 MHz) before applying it to the IQ modulator. Following detection, Bob applies bandpass filtering (0.2 MHz to 1300.0 MHz) to distill the intermediate frequency signals while reducing out-of-band noise. The subsequent downconversion to baseband simultaneously recovers the $x$ and $p$ components. To minimize inter-symbol interference, Alice employs pulse shaping, and Bob utilizes a root-raised-cosine matched filter with a roll-off factor of 0.3 to recover the modulation pulses. For fast-fading phase noise suppression, Bob uses band pass filtering and leveraging the classical pilot symbol to compensate for carrier frequency shift. Quadrature phase-shift keying training symbols are inserted between quantum signals to provide additional phase information to Bob. A phase recovery process is applied to reduce the impact of slow-fading phase noise.

\begin{figure}
    \centering
    \includegraphics[width= 8.5 cm]{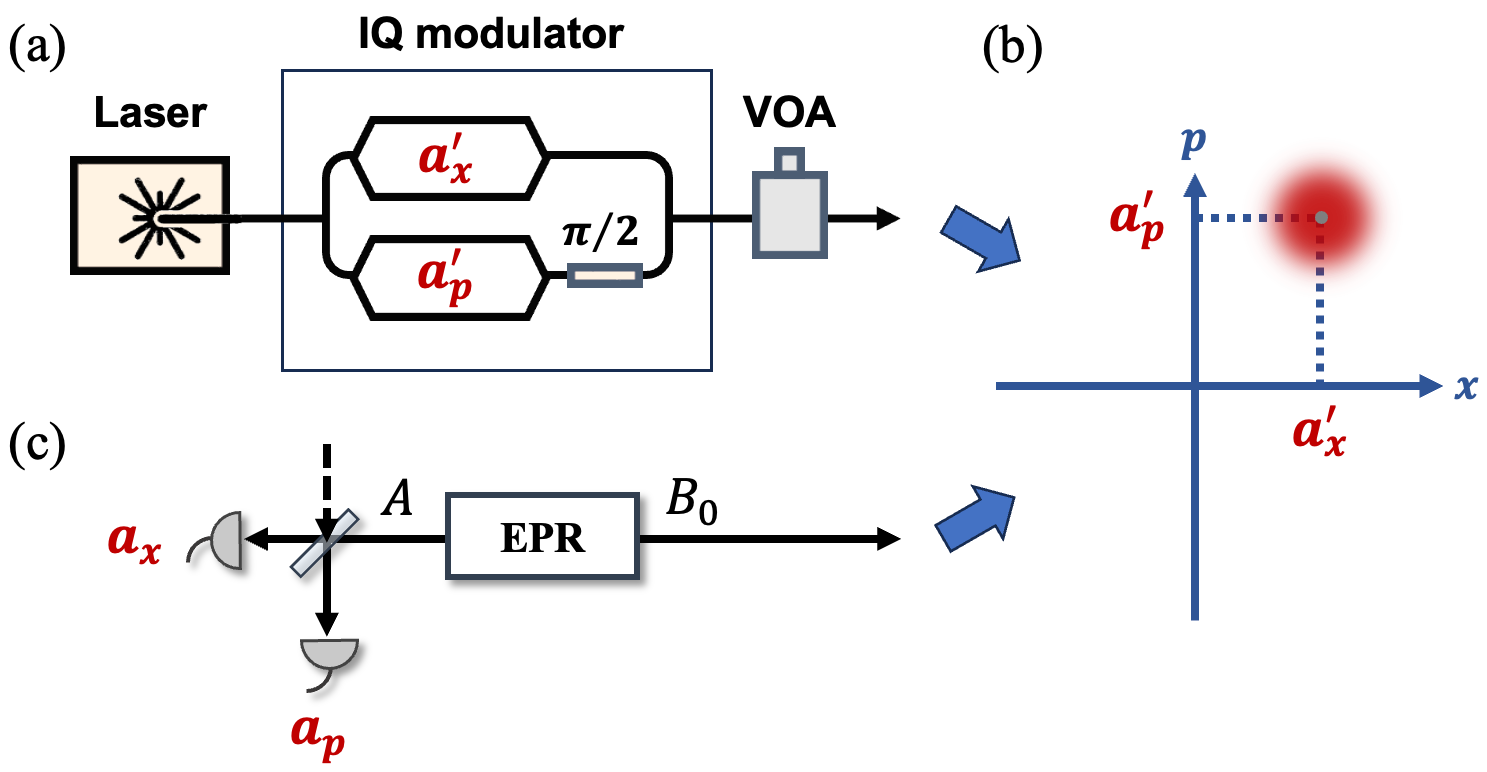}
    \caption{The equivalent source replacement. (a) Preparing a coherent state experimentally with modulation data $(a_x ', a_p ')$. (b) The prepared coherent state in phase space. (c) The equivalent entanglement-based source scheme. }\label{Fig: ExpEqSource}
\end{figure}

\begin{figure}
    \centering
    \includegraphics[width= 4.0 cm]{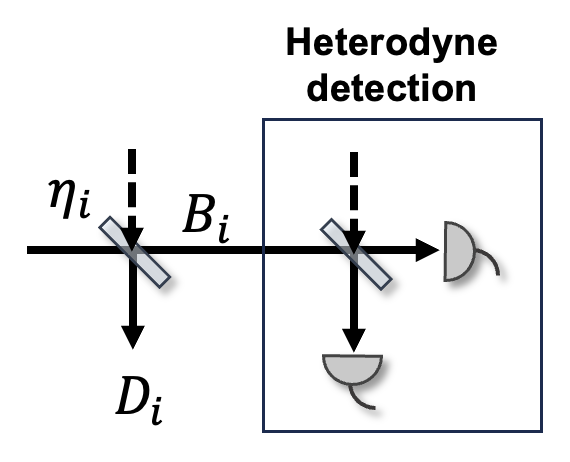}
    \caption{The practical detection scheme. The limited detection efficiency of $\eta_i$ in a practical heterodyne detection is modeled by the beam splitter with transmittance $\eta_i$. The security of this detector model is established with one-time shot noise unit calibration \cite{zhang2020one}. }\label{Fig: Detection}
\end{figure}

\begin{figure}
    \centering
    \includegraphics[width= 7 cm]{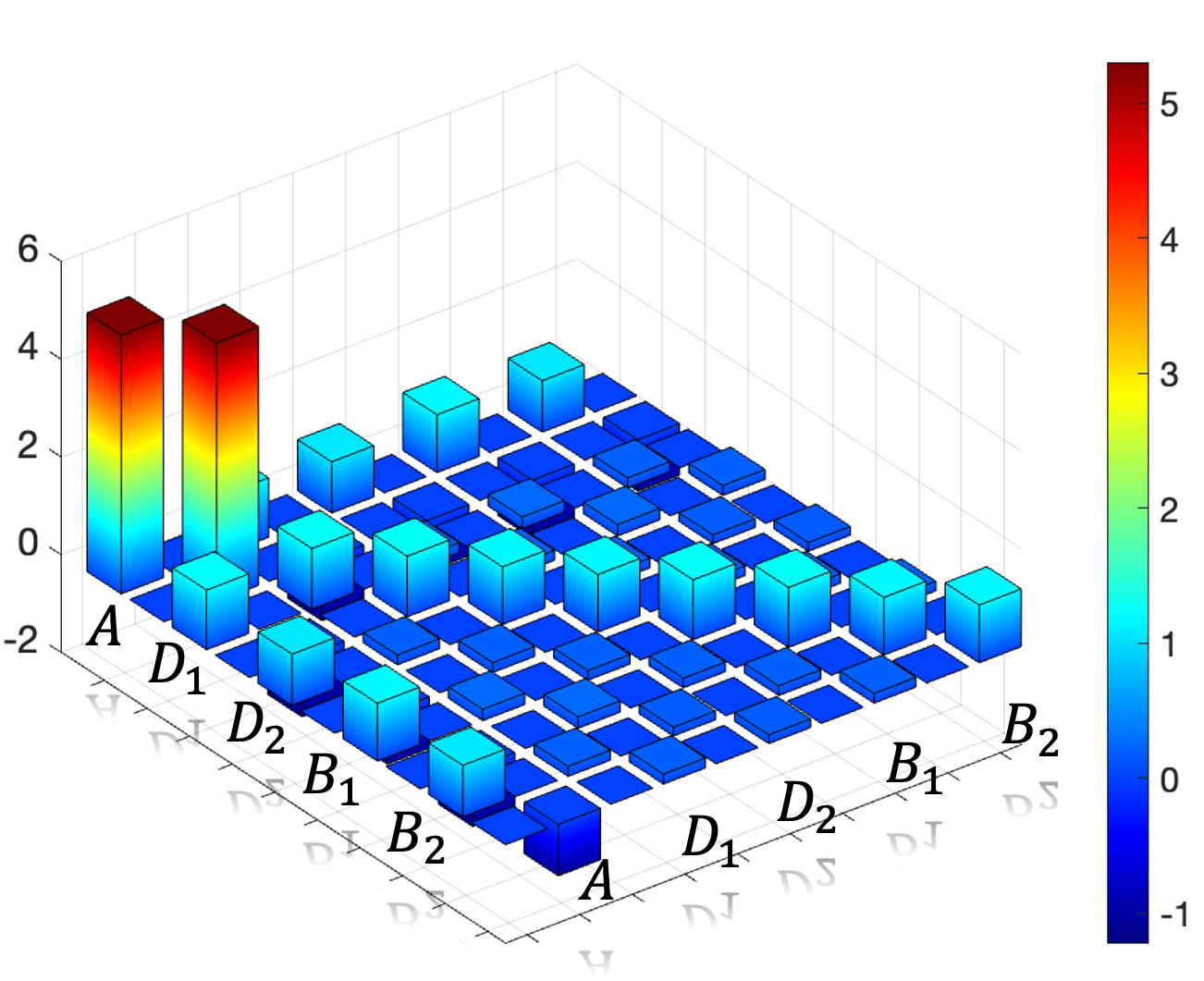}
    \caption{The covariance matrix, $\gamma_{AD_1 D_2 B_1 B_2}$, achieved in one round of our experiment. }\label{Fig: ExpCovMat}
\end{figure}

\subsection{Experimental key rate calculation based on the equivalent entanglement-based scheme}
The experimental key rate is achieved based on the covariance matrix in an entanglement-based scheme that is equivalent to the prepare-and-measure scheme we experimentally implement. The key point is the equivalent source replacement as shown in Fig. \ref{Fig: ExpEqSource}. When Alice uses the experimental configuration as shown in Fig. \ref{Fig: ExpEqSource} (a) to prepare a coherent state located on phase space $(a_x ',a_p ')$, as shown in Fig. \ref{Fig: ExpEqSource} (b), the heterodyne detection results in an equivalent entanglement-based scheme as shown in Fig. \ref{Fig: ExpEqSource} (c) is determined by
\begin{equation}
    (a_x ' , a_p ') = \sqrt{2\frac{V-1}{V+1}}(a_x , a_p).
\end{equation}
Here, $V$ is the variance of mode $A$. It can be easily verified that
\begin{equation}
    V_M = \braket{{a_x '} ^2} = 2\frac{V-1}{V+1} \braket{{a_x} ^2} = V-1,
\end{equation}
with $\braket{{a_x} ^2} = (V+1)/2$. Here, $V_M$ is modulation variance, the key parameter in a prepare-and-measure scheme. After transforming $(a_x ' , a_p ')$ to $(a_x , a_p)$, we can achieve the covariance matrix $\gamma_{AD_1 D_2 B_1 B_2}$ by calculating the covariance with Bob 1's detection results $(b_{1x},b_{1p})$ and Bob 2's detection results $(b_{2x},b_{2p})$, as well as their detection efficiencies. Here, mode $D_1$ and $D_2$ characterize the limited detection efficiency at Bob's site, using the detection model as shown in Fig. \ref{Fig: Detection}.

\begin{figure}
    \centering
    \includegraphics[width= 7 cm]{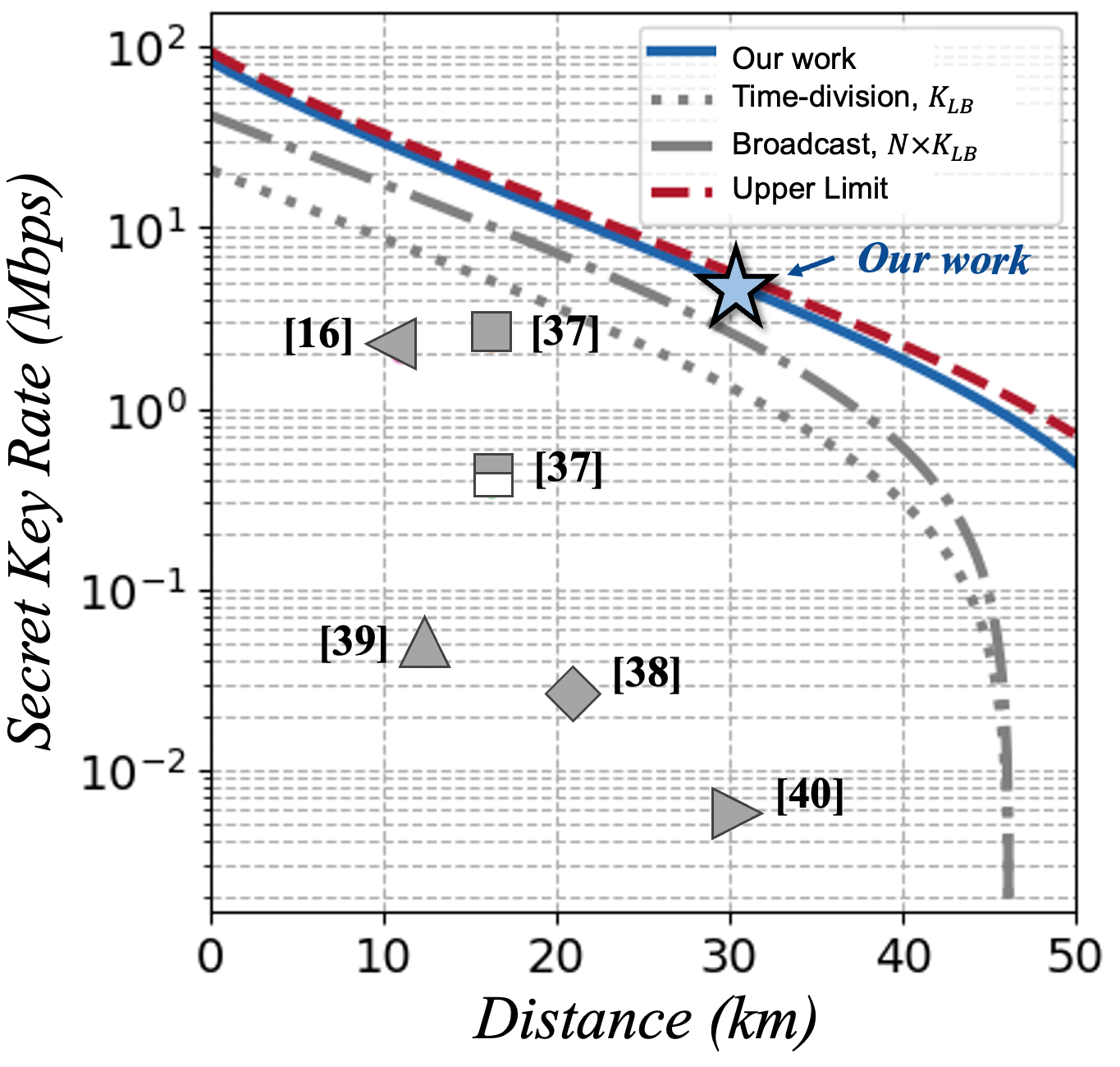}
    \caption{Experimental key rate and simulations. The five-pointed star corresponds to the experimental result of $K_{tot}$ at 30 km. The simulated key rates within 50 km using the same experimental parameters include our protocol (blue solid line), time-division strategy \cite{Huang2021Realizing} (gray dotted line), broadcast strategy \cite{hajomer2024continuous} (gray dash-dotted line) and practical achievable upper limit (red dash line). For comparison, previous state-of-the-art results, including the network using single photons \cite{QNetNature2013,QCnetOE2021} and continuous variables \cite{Huang2020experimental,Xu2023Round,hajomer2024continuous} are included.}\label{Fig: ExpResult}
\end{figure}

The covariance matrix we experimentally achieved is shown in Fig. \ref{Fig: ExpCovMat}. Here, the variance of mode $A$ is 5.3 SNU, indicating a modulation variance of 4.3 SNU and an average photon number of 2.15. The correlations between Alice and any Bob (covariance over 1) are much higher than the correlations between the two Bobs (covariance below 0.25). The simulated $K_{tot}$ with the experimentally achieved parameters is shown in Fig. \ref{Fig: ExpResult} (b). We also simulate the results of $K_{LB}$ and $N \times K_{LB}$, which characterize the key rates in Ref. \cite{Huang2021Realizing} and Ref. \cite{hajomer2024continuous} with the same parameters for a fair comparison. These two methods rely on a pessimistic assumption that Eve fully controls the other users' modes to simplify the security analysis, leading to a limited performance. In contrast, our framework allows for assumptions that more closely reflect the actual operational scenario, i.e., the other users' modes are measured and Eve can optimize her attack with their raw data. Notably, our approach can also accommodate cases in which only a subset of network participants, e.g., Bob 1 to Bob $M$, provides the measurement results during parameter estimation. In this case, Eve's knowledge about Bob 1 can be bounded by $I(b_1 : b_{23...M}) + S(b_1 : E | b_{23...M})$, assuming Eve can purify the system $AB_1 B_2 ... B_M$.

\end{document}